\documentclass[12pt]{article}
\usepackage{amsmath}
\usepackage{multirow}
\usepackage{amsfonts}
\usepackage{amssymb,graphics,psfrag,float}
\usepackage{array,epsfig,multirow,graphicx}
\usepackage{comment}
\usepackage{slashed}
\usepackage{hyperref}
\usepackage{tensor}
\usepackage{booktabs}
\usepackage{colortbl}
\usepackage{hhline}
\usepackage{mathtools}
\usepackage{enumitem}
\usepackage{xfrac}
\usepackage{tikz}
\usepackage[nosort]{cite}
\usetikzlibrary{decorations.markings}
\usetikzlibrary{shapes,arrows}
\usetikzlibrary{decorations.pathreplacing}
\usetikzlibrary{arrows,positioning}

\def\hybrid{\topmargin -20pt    \oddsidemargin 0pt
        \headheight 0pt \headsep 0pt 
        \textwidth 6.25in      
        \textheight 9 in      
        \marginparwidth .875in
        \parskip 5pt plus 1pt
          \jot = 1.5ex
  }
\hybrid
\numberwithin{equation}{section}
\numberwithin{table}{section}

\newcommand{\beq}{\begin{equation}}
\newcommand{\eeq}{\end{equation}}
\newcommand{\be}{\begin{equation}}
\newcommand{\ee}{\end{equation}}
\newcommand{\bea}{\begin{eqnarray}}
\newcommand{\eea}{\end{eqnarray}}
\newcommand{\ben}{\begin{eqnarray*}}
\newcommand{\een}{\end{eqnarray*}}               
\newcommand{\ba}{\begin{aligned}}
\newcommand{\ea}{\end{aligned}}
\newcommand{\bt}{\begin{tabular}}
\newcommand{\et}{\end{tabular}}
\newcommand{\bc}{\begin{center}}
\newcommand{\ec}{\end{center}}

\newcommand{\cS}{\mathcal{S}}
\newcommand{\cK}{\mathcal{K}}
\newcommand{\cN}{\mathcal{N}}

\newcommand{\cF}{\mathcal{F}}

\newcommand{\cV}{\mathcal{V}}

\newcommand{\cM}{\mathcal M}

\newcommand{\I}{\text{Im}}
\newcommand{\R}{\text{Re}}

\newcommand{\bbZ}{\mathbb{Z}}

\newcommand{\bbC}{\mathbb{C}}

\newcommand{\nn}{\nonumber}

\newcommand{\cref}{{\bf [check ref]}}

\definecolor{mppgreen}{RGB}{17,102,86}
\definecolor{mppgray}{RGB}{221,222,214}

\def\blfootnote{\xdef\@thefnmark{}\@footnotetext}
\long\def\symbolfootnote[#1]#2{\begingroup%
\def\thefootnote{\fnsymbol{footnote}}\footnote[#1]{#2}\endgroup}

\begin{document}

\baselineskip=15pt

\begin{titlepage}
\begin{flushright}
\parbox[t]{1.8in}{\begin{flushright}
MPP-2015-278 \end{flushright}}
\end{flushright}

\begin{center}

\vspace*{ 1.2cm}

{\Large \bf  On Mirror Symmetry for Calabi-Yau Fourfolds \\[.3cm]
with Three-Form Cohomology}

\vskip 1.2cm

\renewcommand{\thefootnote}{}
\begin{center}
 {Sebastian Greiner and Thomas W.~Grimm \ \footnote{sgreiner,\ grimm@mpp.mpg.de}}
\end{center}
\vskip .2cm
\renewcommand{\thefootnote}{\arabic{footnote}}

Max-Planck-Institut f\"ur Physik, \\
F\"ohringer Ring 6, 80805 Munich, Germany\\[.1cm]

and\\[.1cm]

Institute for Theoretical Physics and \\
Center for Extreme Matter and Emergent Phenomena,\\
Utrecht University, Leuvenlaan 4, 3584 CE Utrecht, The Netherlands

 \vspace*{.2cm}

\end{center}

 \renewcommand{\thefootnote}{\arabic{footnote}}
 
\begin{center} {\bf ABSTRACT } \end{center}
We study the action of mirror symmetry on 
two-dimensional $\cN=(2,2)$ effective theories obtained 
by compactifying Type IIA string theory on Calabi-Yau fourfolds. 
Our focus is on  
fourfold geometries with non-trivial three-form cohomology. The 
couplings of the massless zero-modes arising by expanding in these forms
 depend both on the complex structure deformations and 
the K\"ahler structure deformations of the Calabi-Yau 
fourfold. We argue that two holomorphic functions of the deformation 
moduli capture this information. These are exchanged under mirror symmetry,
which allows us to derive them at the large complex structure and large volume 
 point. We discuss the application of the resulting explicit expression 
 to F-theory compactifications and their weak string coupling limit. 
 In the latter orientifold settings we demonstrate 
compatibility with mirror symmetry of Calabi-Yau threefolds at large complex structure. 
As a byproduct we find an interesting relation of no-scale like conditions 
on K\"ahler potentials to the existence of chiral and twisted-chiral descriptions in 
two dimensions.

\hfill {December, 2015}
\end{titlepage}

\tableofcontents

\newpage



\section{Introduction}

The derivation of four-dimensional 
low-energy effective actions arising from string theory requires 
a detailed understanding of the geometries used as compactification 
spaces. Since the early days of string theory much research has focused on the study of 
Calabi-Yau manifolds of complex dimension three. These threefolds 
were identified as valid compactification backgrounds to four space-time 
dimensions and can yield to, for example when used in  
the heterotic string theories, potentially 
phenomenologically interesting four-dimensional effective theories
with the minimal amount of supersymmetry. In contrast, there is 
significantly less known about the geometry of 
Calabi-Yau manifolds of complex dimension four. 
With the advent of F-theory \cite{Vafa:1996xn,Denef:2008wq,Weigand:2010wm} it became clear that 
these fourfolds are relevant in obtaining 
four-dimensional effective theories
with the minimal amount of supersymmetry from Type IIB
string theory. It is therefore crucial to further our understanding 
of the geometry of Calabi-Yau fourfolds and investigate
the relation to couplings in the effective theories. 

In contrast to Calabi-Yau threeforlds one finds that 
Calabi-Yau fourfolds admit three non-trivial independent 
Hodge numbers that count the number of 
harmonic forms of different degree. In full analogy to threefolds,
two of them encode the number of complex structure and 
K\"ahler structure deformations of the geometry. 
When compactifying string theory or M-theory 
on a Calabi-Yau fourfold, these deformations will appear 
as massless fields, so-called moduli, in the effective action. 
The moduli space geometry has been studied in various 
works \cite{Greene:1993vm,Mayr:1996sh,Klemm:1996ts,Grimm:2009ef,
Honma:2013hma,Halverson:2013qca,Braun:2014xka,Bizet:2014uua}. The additional Hodge number on 
Calabi-Yau fourfolds is associated to the number 
of harmonic three-forms. In this work we will discuss in detail 
how the presence of these three-forms affects the effective theory 
when compactifying Type IIA string theory and
M-theory on Calabi-Yau fourfolds. The effective theory 
will then admit new scalars $N_l$ with couplings non-trivially 
varying over both the complex structure and K\"ahler structure 
moduli spaces. This was already observed for M-theory compactifications 
in \cite{Haack:1999zv,Haack:2001jz}, for Type IIA compactifications in \cite{Haack:2000di}, 
and for F-theory compactifications in \cite{Grimm:2010ks}.
We will show in this work that the moduli dependence at certain 
points in the moduli space can actually be computed explicitly by 
using mirror symmetry for Calabi-Yau fourfolds.  

Our first focus is on a refined understanding 
of the moduli variations of three-forms. Therefore, we begin 
by introducing a basis 
of $(2,1)$-forms on the fourfold that is convenient when performing 
the dimensional reduction. The so-defined set of forms is adapted to the 
underlying complex structure and it was pointed out in \cite{Grimm:2010ks}
that their variation with the complex structure moduli can 
be captured by a holomorphic function $f_{kl}$, with indices 
ranging over the number of harmonic $(2,1)$-forms. 
This holomorphic function can be used in the compactification 
of Type IIA string theory, accessed via its low-energy supergravity theory, 
on a Calabi-Yau fourfold. Such dimensional reductions
of the Type IIA theory have already been investigated in \cite{Gukov:1999ya,Haack:2000di,Gates:2000fj}.
They are expected to yield  two-dimensional effective theories with $\cN=(2,2)$ supersymmetry 
describing the dynamics of chiral and twisted-chiral multiplets.  
Without including three-forms the supersymmetry properties of such Type IIA effective theories 
were already discussed in \cite{Gates:2000fj} by extending earlier 
results \cite{Gates:1994gx,deWit:1992xr,Grisaru:1995dr}. We will consider the generalization 
of this result including the three-form scalars $N_l$ 
and suggest that it leads to a more general class of supersymmetric 
dilaton supergravities. 

The two-dimensional $\cN=(2,2)$ effective action is expected to be 
invariant under the action of mirror symmetry. More precisely, considering 
Type IIA string theory on two Calabi-Yau fourfolds that are mirror manifolds to each other,
the resulting two effective actions should admit an identification under 
an appropriate mirror map. This map identifies complex coordinates and couplings 
at special points in moduli space.
Mirror symmetry exchanges complex structure and K\"ahler structure deformations, but 
preserves the number of non-trivial three-forms and thus the number of three-form scalars $N_l$. 
It also maps chiral to twisted-chiral 
multiplets. Therefore, we are forced to perform an appropriate duality transformation 
for the three-form scalars appearing in pairs of effective actions arising 
from mirror manifolds. In both 
effective actions the dynamics of the three-forms are described by two
holomorphic functions $f_{kl}$ and $h_l^k$. The former is holomorphic
in the complex structure moduli, while the latter is holomorphic in
the complexified 
K\"ahler moduli. These functions are exchanged by mirror symmetry 
and we are able to derive the complex structure dependence of $f_{kl}$ in the 
large complex structure limit by using the results of a large volume compactification
on the mirror geometry. 
 
Our results have several interesting applications, in particular, when using 
the Calabi-Yau fourfolds with non-trivial three-forms 
as F-theory backgrounds. To determine the four-dimensional 
F-theory effective actions for such configurations one uses the duality with 
M-theory \cite{Vafa:1996xn,Denef:2008wq,Grimm:2010ks}.
It was shown in \cite{Grimm:2010ks}  
that the function $f_{lm}$ can either lift to a gauge coupling function 
of R-R vector fields or to the metric of a special set of complex scalars. 
In both cases it is desirable to explicitly 
compute the moduli dependence of their coupling function.
For example, the three-form scalars lifting to four-dimensional 
scalars naturally admit real shift symmetries or even a generalized 
Heisenberg symmetry and might be of profound phenomenological interest
(see, for example, \cite{Grimm:2011tb,Grimm:2014vva,Grimm:2015ona}). Furthermore,
considering the weak string coupling limit of the F-theory setting following \cite{Sen:1996vd,Sen:1997gv} 
the resulting effective theory should match with the orientifold effective 
actions \cite{Grimm:2004uq,Grimm:2004ua}. In the case that such a limit exists one can associate 
a Calabi-Yau threefold to the F-theory Calabi-Yau fourfold. We 
are then able to show that our result for $f_{kl}$ obtained by 
fourfold mirror symmetry is consistent with the weak coupling 
analog obtained from threefold mirror symmetry.

This paper is organized as follows. In \autoref{three-form_facts} we recall some 
basics about Calabi-Yau fourfolds and discuss a convenient basis of $(2,1)$-forms 
parametrized by a holomorphic function $f_{lk}$. We dimensionally reduce 
Type IIA supergravity on a Calabi-Yau fourfold in \autoref{IIAreduction}.
This allows us to investigate the $\cN=(2,2)$ supersymmetric structure 
of the effective theory and perform a set of important dualizations to 
interchange chiral multiplets and twisted-chiral multiplets. In \autoref{mirror_section}
we discuss mirror symmetry with a focus on the $(2,1)$-form sector. 
This allows us to determine $f_{lm}$ in the large complex structure limit. 
We use these results in an F-theory compactification on 
an elliptically fibered Calabi-Yau fourfold in \autoref{F-theoryapp}. 
Moving to the weak string coupling limit, we find compatibility of 
our result for $f_{lm}$ with the answers predicted by mirror symmetry 
of Calabi-Yau threefolds. This work has two appendices with useful 
computational results. In \autoref{3d-2dreduction} we perform the  
circle reduction of a general  three-dimensional un-gauged $\cN=2$ supergravity theory
with focus on the bosonic action. We find interesting
conditions on the kinetic potential to match the proposed $\cN=(2,2)$ action in two dimensions. 
The dualization of chiral to twisted-chiral multiplets in the bosonic sector is 
performed in \autoref{detailed_dual}. We again find conditions on the kinetic potential
in order that this dualization can be performed. The results of both 
appendices are immediately applicable to Calabi-Yau fourfold effective actions 
of Type IIA string theory and M-theory.

\section{On the geometry of Calabi-Yau fourfolds with three-form cohomology} \label{three-form_facts}

In this section we introduce important facts about the geometry of Calabi-Yau fourfolds $Y_4$. A brief summary 
of some basics about their differential structure and topology will be given in \autoref{CY4basics}.
The focus of \autoref{3formbasics} will be to introduce relevant properties of the three-form 
cohomology of $Y_4$. We argue that an appropriate definition of three-forms of Hodge-type 
(2,1) can be given in terms of a function $f_{mn}$ holomorphic in the complex structure moduli.
This function will be of key interest throughout this work. 

\subsection{Some basic properties of Calabi-Yau fourfolds} \label{CY4basics}

We define a compact real eight-dimensional manifold $Y_4$ to be a Calabi-Yau fourfold if its holonomy group is 
exactly $SU(4)$. Such manifolds are K\"ahler, i.e.~admit a closed K\"ahler two-form $J$, and possess a unique Ricci-flat metric within the class of $J$. Furthermore, one can introduce a non-trivial closed $(4,0)$-form $\Omega$ on $Y_4$ that is unique up to constant rescalings. Note that $J$ and $\Omega$ can be used to form a 
top-form on $Y_4$ and one has
\beq \label{def-Omega2}
   \frac{1}{4!} J \wedge J \wedge J \wedge J = |\Omega|^{-2}\, \Omega \wedge \bar \Omega \ , \qquad \qquad  |\Omega|^2= \frac{1}{\cV} \int_{Y_4} \Omega \wedge \bar \Omega \ ,
\eeq
where $\cV$ is the total volume of $Y_4$. The $SU(4)$ holonomy also allows one to introduce
one complex covariantly constant and no-where vanishing spinor of definite chirality. 
The forms $J$ and $\Omega$ are obtained as bilinear contractions using this spinor.

 With our definition of a Calabi-Yau fourfold, we can also constrain  the Hodge numbers 
 $ h^{p,q} (Y_4) = \text{dim}(H^{p,q}(Y_4,\bbC) )$. 
 There are three independent Hodge numbers on $Y_4$: $h^{1,1}(Y_4) $, $h^{3,1}(Y_4) $, and $h^{2,1}(Y_4) $.  
The significance of $h^{1,1}(Y_4) $ and $h^{3,1}(Y_4) $ in the dimensional reduction are very similar to 
the case of a Calabi-Yau threefold (see e.g.~\cite{Candelas:1990pi}).  On the one hand, the
 number $ h^{1,1} (Y_4)$ counts the allowed K\"ahler structure deformations, which we will 
denote by $v^A$. On the other hand, the number 
$ h^{3,1} (Y_4)$ counts the complex structure deformations denoted by $z^K$. Both 
turn out to become moduli fields in the effective theory obtained by dimensional reduction on $Y_4$
and will be discussed in more detail in \autoref{IIAreduction_details}. 
The Hodge number $h^{2,1}$ has no threefold analog and understanding the geometries 
with $h^{2,1}(Y_4)>0$ will be the main focus of this work.
Having three independent Hodge numbers, the Hodge diamond takes the form 

\vspace{.5cm}
\arraycolsep=2,0pt\def\arraystretch{1.4}
\setlength{\unitlength}{0.6cm}
\begin{picture}(21,8)
 \put(5,8){$ h^{0,0} $}
 \put(4,7){$ h^{1,0} $} \put(6,7){$ h^{0,1} $}
 \put(3,6){$ h^{2,0} $} \put(5,6){$ h^{1,1} $} \put(7,6){$ h^{0,2} $}
 \put(2,5){$ h^{3,0} $} \put(4,5){$ h^{2,1} $} \put(6,5){$ h^{1,2} $} \put(8,5){$ h^{0,3} $}
 \put(1,4){$ h^{4,0} $} \put(3,4){$ h^{3,1} $} \put(5,4){$ h^{2,2} $} \put(7,4){$ h^{1,3} $} \put(9,4){$ h^{0,4} $}
 \put(2,3){$ h^{4,1} $} \put(4,3){$ h^{3,2} $} \put(6,3){$ h^{2,3} $} \put(8,3){$ h^{1,4} $}
 \put(3,2){$ h^{4,2} $} \put(5,2){$ h^{3,3} $} \put(7,2){$ h^{2,4} $}
 \put(4,1){$ h^{4,3} $} \put(6,1){$ h^{3,4} $}
 \put(5,0){$ h^{4,4} $}
 
 \put(11,4){$ = $}
 
 \put(17,8){$ 1 $}
 \put(16,7){$ 0 $} \put(18,7){$ 0 $}
 \put(15,6){$ 0 $} \put(17,6){$ h^{1,1} $} \put(19,6){$ 0 $}
 \put(14,5){$ 0 $} \put(16,5){$ h^{2,1} $} \put(18,5){$ h^{2,1} $} \put(20,5){$ 0 $}
 \put(13,4){$ 1 $} \put(15,4){$ h^{3,1} $} \put(17,4){$ h^{2,2} $} \put(19,4){$ h^{3,1} $} \put(21,4){$ 1\ \ , $}
 \put(14,3){$ 0 $} \put(16,3){$ h^{2,1} $} \put(18,3){$ h^{2,1} $} \put(20,3){$ 0 $}
 \put(15,2){$ 0 $} \put(17,2){$ h^{1,1} $} \put(19,2){$ 0 $}
 \put(16,1){$ 0 $} \put(18,1){$ 0 $}
 \put(17,0){$ 1 $}
 \setlength{\unitlength}{0.06cm}
 \multiput(20,73)(10,-10){7}{\line(1,-1){6}}       
\end{picture}

\noindent
where we have indicated for later use the action of mirror symmetry on the Hodge numbers. More
precisely, mirror symmetry identifies two Calabi-Yau geometries with Hodge numbers 
mirrored along the dashed line. 
A more detailed discussion of mirror symmetry will be presented in \autoref{mirror_section}. 
In addition, one finds the formulas \cite{Klemm:1996ts}
\beq
 h^{2,2}(Y_4) = 2(22 + 2h^{1,1} + 2h^{3,1} - h^{2,1})\ , \qquad  \quad \chi(Y_4) 
 = 6(8 + h^{1,1} + h^{3,1} - h^{2,1})
\eeq
where  $ \chi = \sum_{p,q} (-1)^{p+q} h^{p,q}$ is the Euler characteristic of $Y_4$.

\subsection{Non-trivial three-forms on Calabi-Yau fourfolds} \label{3formbasics}

Of key importance in this work is the inclusion of non-trivial three-forms in the dimensional reduction and discussion 
of mirror symmetry. In this subsection we summarize some basic properties of such three-forms that 
will be useful throughout the later sections. 

To begin with, we comment on the moduli dependence of three-forms when 
choosing them to represent elements of $H^{2,1}(Y_4)$. 
In order to do that, recall that the Hodge filtration of the three-cohomology 
$ H^3 (Y_4, \mathbb{C} ) $ is given by the holomorphic bundles 
$ F^p (Y_4) = \bigoplus_{j=p} ^3 H^{j, 3-j} $ over the complex structure moduli space. 
Since $ H^{3,0}(Y_4) $ is trivial, this enables us to find a basis $ \psi_l $ of 
$F^2(Y_4) = H^{2,1}(Y_4) $, which varies holomorphically with the complex structure moduli $z^K$, i.e.~one has
\beq
\frac{\partial}{ \partial {\bar{z}^K} }\psi_l = 0, \qquad l=1, \ldots, h^{2,1}(Y_4)\ ,  
\eeq
where $K = 1, \ldots, h^{3,1}(Y_4)$ labels the complex structure moduli. 
Note that in the dimensional reduction we can think of $\psi_l$ to be 
the harmonic representative in each class of $H^{2,1}(Y_4) $.
At a given point in the complex structure space we can write this basis in the form
\beq \label{def-f1}
\psi_l = \alpha_l + i f_{lm} (z) \beta^m   \qquad \in H^{2,1}(Y_4)\ ,
\eeq
where $ (\alpha_l ,\beta^m )$ comprise a real moduli-independent basis of $ H^3(Y_4, \mathbb{R}) $.\footnote{
It might be natural to chose $( \alpha_l , \beta^m)$ to be a basis of $ H^3(Y_4, \mathbb{Z})$, but quantization 
of the coefficients will not be important in this work.}
Holomorphicity of the forms $\psi_l$ translates to the fact that 
 $ f_{lm}(z) $ is a holomorphic function of the complex structure moduli $ z^K$. 
 Furthermore, we assume that $(\alpha_l,\beta^m)$ is chosen such 
 that $\R f_{lm}$ is a positive definite and invertible matrix.\footnote{While we have no 
 complete proof that this is always possible, we note that $H^{2,1}(Y_4)/H^{3}(Y_4,\bbZ)$ is 
 actually a complex torus. $f_{lm}$ sets the complex structure on this torus.}

After performing the dimensional reduction on $Y_4$ in \autoref{IIAreduction}, we 
aim to find the proper complex fields that are compatible with two-dimensional supersymmetry. 
For the zero-modes arising from the $\psi_l$ it turns out that a further 
normalization is useful, i.e.~we introduce the (1,2)-forms
\beq \label{def-Psil}
  \Psi^l = \frac{1}{2} (\R f)^{lm} \bar {\psi}_m =  \frac{1}{2} (\R f)^{lm} ( \alpha_m - i \bar f_{mn} (z) \beta^n) 
  \qquad \in H^{1,2}(Y_4)\ .
\eeq
In this expression we have multiplied by the \textit{inverse} $(\R f)^{lm}$ of the real part of $ f_{lm}$, i.e.
$ (\R f)^{lm} \, (\R f)_{mk}= \delta^l _k $.
This definition allows to give a simple expressions for $\I\, \Psi^l$ and the derivative 
of $ \Psi^l$ with respect to the complex structure moduli:
\beq \label{useful}
  \bar \Psi^l - \Psi^l = i \beta^l\ ,\qquad  \partial_{z^K} \Psi^l = - \Psi^k \, (\R f)^{lm}\, \partial_{z^K} (\R f)_{mk}\ ,
\eeq
and accordingly $ \partial_{z^K} \Psi^l =  \partial_{z^K} \bar \Psi^l$.
Note that $\Psi^l$ is a $(1,2)$-form and therefore satisfies
\beq \label{starPsi}
    * \Psi^l =- i J \wedge \Psi^l\ ,
\eeq
where $*$ is the Hodge-star for the Calabi-Yau metric on $Y_4$. 

To evaluate the integrals appearing in the dimensional reduction we
impose one further condition on the basis $(\alpha_l,\beta^m)$. 
More precisely, we demand 
\beq
\beta^l \wedge \beta^m = 0\ , \qquad \forall \  l,m = 1, \ldots, h^{2,1}(Y_4)\ ,
\eeq
which is supposed to hold in cohomology.\footnote{ 
Considering hypersurfaces in toric varieties, this condition can be satisfied for non-trivial three-forms
arising from singular Riemann surfaces. This allows us to choose a symplectic basis 
with respect to a certain divisor for the three-forms.}
Introducing a basis $\omega_A$ of $H^{1,1}(Y_4)$ we thus find that 
\beq \label{betabeta_condition}
    \int_{Y_4} \omega_A \wedge \beta^l \wedge \beta^m = 0 \ ,\qquad \forall \  A = 1, \ldots, h^{1,1}(Y_4) \ .
\eeq
The remaining integrals are in general non-trivial and denoted 
by  
\beq \label{Cdef}
 C_{Am}{}^k = \int_{Y_4} \omega_A \wedge \alpha_m \wedge \beta^k\ ,\qquad 
  C_{Amk} = \int_{Y_4} \omega_A \wedge \alpha_m \wedge \alpha_k\ .
\eeq
Using a basis  $(\alpha_m, \beta^k)$ satisfying \eqref{betabeta_condition}
one checks that the metric $\int \Psi^l \wedge * \bar \Psi^k$ is symmetric in the 
indices $l,k$ and real. This property will be crucial in determining a 
kinetic potential for this metric.

\section{Dimensional reduction of Type IIA supergravity}  \label{IIAreduction}

In this section we perform the dimensional reduction of Type IIA supergravity on 
a Calabi-Yau fourfold $Y_4$. Such reductions have  already been
performed in \cite{Haack:1999zv,Haack:2001jz,Haack:2000di,Gates:2000fj}. Our analysis follows 
\cite{Haack:1999zv,Haack:2001jz,Haack:2000di}, but we will apply in addition the improved understanding about 
the three-form cohomology of \autoref{three-form_facts}.

\subsection{The effective action from a Calabi-Yau fourfold reduction} \label{IIAreduction_details}

The Kaluza-Klein reduction of Type IIA supergravity can be trusted in the
limit in which the typical length scale of the physical volumes of submanifolds of $Y_4$ are sufficiently 
large compared to the string scale. This limit is referred to 
as the large volume limit. Furthermore, these typical length scales 
set the Kaluza-Klein scale which we assume to be sufficiently above the 
energy scale of the effective action. We therefore keep only the massless Kaluza-Klein modes 
in the following reduction. 

Our starting point will be the bosonic part of the ten-dimensional Type IIA action in string-frame given by \footnote{
Note that for convenience we have set $\kappa^2 = 1$.}
\begin{align} \label{10Daction}
S^{(10)}_{\rm IIA} &= \int e^{-2\check \phi_{\rm IIA}} \left( \frac{1}{2}\check R \, \check \ast 1 + 2 d\check \phi_{\rm IIA} \wedge \check \ast d\check \phi_{\rm IIA} - \frac{1}{4} \check H_3 \wedge \check \ast \check  H_3 \right) \nonumber \\
									&\quad - \frac{1}{4} \int \Big( \check F_2 \wedge \check \ast \check F_2 + \check {\mathbf F}_4 \wedge \check \ast \check {\mathbf{F}}_4+ \check B_2 \wedge \check F_4 \wedge \check F_4\Big) \ .
\end{align}
where $\check \phi_{\rm IIA}$ is the ten-dimensional dilaton, $ \check H_3 = d \check B_2 $ is the field strength of the NS-NS two-form 
$ \check B_2 $, and $\check F_p = d\check C_p$ are the field strengths of the R-R $p$-forms $\check C_1$ and $\check C_3$. 
We also have used the modified field strength $ \check {\mathbf{F}}_4 = \check F_4 - \check C_1 \wedge \check H_3 $. 
Here and in the following we will use a check to indicate ten-dimensional fields. 

The background solution around which we want to consider the effective theory is 
taken to be of the form $\mathbb{M}_{1,1} \times Y_4$, where $\mathbb{M}_{1,1}$
is the two-dimensional Minkowski space-time, and $Y_4$ is a Calabi-Yau fourfold 
with properties introduced in \autoref{three-form_facts}. As pointed out there 
such a manifold admits one complex covariantly constant spinor of definite chirality. 
This spinor can be used to dimensionally reduce the $\cN=(1,1)$ supersymmetry 
of Type IIA supergravity to obtain a two-dimensional $\cN=(2,2)$ supergravity theory.
In particular, the two ten-dimensional gravitinos of opposite chirality reduce to two 
pairs of two-dimensional Majorana-Weyl gravitinos with opposite chirality. We will have more to 
say about the supersymmetry properties of the two-dimensional action in \autoref{22dilatonSugra}.
Furthermore, recall that $Y_4$ admits a 
Ricci-flat metric $g^{(8)}_{mn}$ and one can thus check that a metric of the form 
\beq  \label{metric-ansatz}
   d\check s^2 = \eta_{\mu \nu} dx^\mu dx^\nu + g^{(8)}_{mn} dy^m dy^n\ , 
\eeq
solves the ten-dimensional equations of motion in the absence of background fluxes.\footnote{The 
inclusion of background fluxes complicates the reduction further. In particular, it 
requires to introduce a warp-factor. The M-theory reduction with warp-factor was recently 
performed in \cite{GPW}.}
Note that in \eqref{metric-ansatz} we denote 
by $ x^\mu $ the two-dimensional coordinates of the space-time $\mathbb{M}_{1,1} $, 
whereas the eight-dimensional real coordinates on the Calabi-Yau fourfold $ Y_4 $ are denoted 
by $ y^m $. 

The massless perturbations around this background both consist of fluctuations 
of the internal metric $g^{(8)}_{mn}$ that preserve the Calabi-Yau condition as
well as the fluctuations of the form fields $\check B_2$, $\check C_1,\check C_3$ and the dilaton $\check \phi_{\rm IIA}$.  
The metric fluctuations give rise to the real K\"ahler structure moduli $v^A$, $A=1,\ldots, h^{1,1}(Y_4)$ that preserve the 
complex structure and are given by 
\begin{equation} \label{Kaehlermoduli}
	g_{i\bar \jmath } + \delta g_{i \bar \jmath} = -i J_{i \bar \jmath} = -i v^A\, (\omega_{A})_{ i \bar \jmath},
\end{equation}
where $J$ is the K\"ahler form on $Y_4$ and $ \omega_A$ comprises a real basis of 
harmonic $(1,1)$-forms spanning $H^{1,1}(Y_4) $. The K\"ahler structure moduli 
appear also in the expression of the total string-frame volume $\cV$ of $Y_4$ given by 
\beq
    \cV \equiv \int_{Y_4} *1  = \frac{1}{4!} \int_{Y_4} J \wedge J \wedge J \wedge J\ .
\eeq
In addition to the K\"ahler structure moduli one finds a set of complex structure moduli $z^K$, $K = 1,\ldots, h^{3,1}(Y_4)$.
These fields parameterize the change in the complex structure of $Y_4$ preserving the class 
of its K\"ahler form $J$. Infinitesimally they are given by the fluctuations $\delta z^K$ as 
\begin{equation} \label{CSmoduli}
	\delta g_{\bar \imath \bar \jmath} = - \frac{1}{3 | \Omega| ^2 } \overline{\Omega}_{\bar \imath}^{\ lmn} (\chi_{K})_{lmn \bar \jmath}\, \delta z^{K} \ ,\qquad  \qquad 
\end{equation}
where $\Omega$ is the $(4,0)$-form, the $\chi_{K}$ form a 
basis of harmonic $(3,1)$-forms spanning $H^{3,1}(Y_4)$, and $|\Omega|^2$
was already given  in \eqref{def-Omega2}. 

The Kaluza-Klein ansatz for the remaining fields takes the form 
\begin{align} \label{BC-expand}
   &\check B_2 = b^A \omega_A \ , \qquad \check C_1 = A \ , \\
   &\check C_3 = V^A \wedge \omega_A + N_l \Psi^l + \bar N_l \bar \Psi^l \ , \nn 
\end{align}
where $\Psi^l$ is a basis of harmonic 
$(1,2)$-forms spanning $H^{1,2}(Y_4)$ as introduced in \eqref{def-Psil}. 
A discussion of the properties of $\Psi^l$ was already given in \autoref{three-form_facts}.
Finally, we dimensionally reduce the Type IIA dilaton by dropping its 
dependence on the internal manifold $Y_4$. It turns out to be convenient 
to define a two-dimensional dilaton $\phi_{\rm IIA}$ in 
terms of the ten-dimensional dilaton $\check \phi_{\rm IIA}$ as
\beq \label{def-2dilaton}
   e^{2 \phi_{\rm IIA}}  \equiv  \frac{e^{2 \check \phi_{\rm IIA}}}{\cV}\ .
\eeq   
In summary, we find in the two-dimensional $\cN=(2,2)$ supergravity theory 
the $2 h^{1,1}(Y_4) + 1$ real scalar fields $v^A(x)$, $b^A(x)$, $\phi_{\rm IIA}(x)$ as 
well as the $h^{3,1}(Y_4) + h^{2,1}(Y_4)$ complex scalar fields $z^K$, $N_l$. 
In addition there are $h^{1,1}(Y_4) + 1$ vectors $A$, $V^A$, which carry, however, 
no physical degrees of freedom in a two-dimensional theory if they are not involved 
in any gauging. Since the effective action considered here contains 
no gaugings, we will drop these in the following analysis. 

To perform the dimensional reduction one inserts the expansions \eqref{Kaehlermoduli}, 
\eqref{CSmoduli}, \eqref{BC-expand}, and \eqref{def-2dilaton} into
 the Type IIA action \eqref{10Daction}.
 It reduces 
to the two-dimensional action 
\begin{align} \label{2Daction}
  S^{(2)} = \int  e^{-2 \phi_{\rm IIA}}& \Big( \frac{1}{2} R\, *1 + 2 d \phi_{\rm IIA} \wedge * \phi_{\rm IIA}
  - G_{K\bar L}\, dz^K \wedge * d\bar z^L - G_{A  B} \, d t^A \wedge * d\bar t^B \Big) \nn\\
  &  -   \frac{1}{2} v^A d_{A}{}^{l  k} D N_l \wedge * \overline{D N}_k 
                                                - \frac{i}{4} d_{A}{}^{l  k}  db^A \wedge (N_l \overline{DN}_k - DN_{l} \bar N_k)  \ .
\end{align}
We note that the NS-NS part, which 
is summarized in the first line of \eqref{10Daction}, reduces to the first line of \eqref{2Daction}, 
while the R-R part, i.e.~the second line of  \eqref{10Daction}, reduces to the second line of \eqref{2Daction}.

Let us introduce the various objects appearing in the action \eqref{2Daction}. 
First, we have defined the complex coordinates 
\beq
    t^A \equiv b^A + i v^A\ ,
\eeq
which combine the K\"ahler structure moduli with the B-field moduli. 
Furthermore, we have introduced the metric \footnote{The second 
equality follows from the cohomological identity $* \omega_A = -\frac{1}{2} \omega_A \wedge J \wedge J + \frac{1}{36} \cV^{-1}\cK_A J \wedge J \wedge J$.}
\beq \label{GABmetric}
    G_{A  B} = \frac{1}{4 \cV } \int_{Y_4} \omega_A \wedge * \omega_B 
    = - \frac{1}{8 \cV} \Big( \cK_{AB} - \frac{1}{18 \cV} \cK_A \cK_B \Big)\ ,
\eeq
where $\cV$, $\cK_A$ and $\cK_{AB}$ are given in terms of the 
 quadruple intersection numbers $\cK_{ABCD}$ as
 \begin{align}
   &\cK_{ABCD} = \int_{Y_4} \omega_A \wedge \omega_B\wedge \omega_C \wedge \omega_D\ , \\
    &\cV =\frac{1}{4!} \cK_{ABCD} v^A v^B v^C v^D\ ,\quad \cK_A =  \cK_{ABCD}   v^B v^C v^D\ , \quad 
    \cK_{AB} =  \cK_{ABCD}   v^C v^D\ . \nn 
 \end{align}
With these definitions at hand, we can further evaluate the metric 
$G_{A  B}$ and show that it can be obtained from a K\"ahler potential as
\beq
    G_{A  B} =- \partial_{t^A} \partial_{\bar t^B} \log \, \cV  \ .
\eeq
Also the metric $G_{K\bar L}$ is actually a K\"ahler metric. It only depends on the complex structure 
moduli $z^K$ and takes the form 
\beq
    G_{K\bar L} = - \frac{\int_{Y_4} \chi_K \wedge \bar \chi_L}{\int_{Y_4} \Omega \wedge \bar \Omega} 
    = - \partial_{z^K} \partial_{\bar z^L} \log \int_{Y_4} \Omega \wedge \bar \Omega\ .
\eeq
Note that both $G_{A  B}$ and $G_{K\bar L}$ are actually positive definite and therefore 
define physical kinetic terms in \eqref{2Daction}. Both terms scale
with the dilaton $\phi_{\rm IIA}$ and it is easy to check that this dependence cannot be 
removed using a Weyl-rescaling of the two-dimensional metric. We will show in 
\autoref{22dilatonSugra} that this is consistent with the form of the $\cN=(2,2)$ dilaton 
supergravity. 

Let us now turn to the R-R part of the action \eqref{10Daction} and 
discuss the couplings appearing in the second line of \eqref{2Daction}.
First, we introduce the coupling function
\beq \label{evaluate_d}
    d_A{}^{lm} \equiv  i \int_{Y_4} \omega_A \wedge \Psi^{l} \wedge \bar \Psi^m 
    = - \int_{Y_4} \omega_A \wedge \Psi^{l} \wedge \beta^m = - \frac{1}{2} (\R f)^{ln}C_{An}{}^m  \ ,
\eeq
where we have used \eqref{useful} to evaluate the second equality, and \eqref{betabeta_condition}, \eqref{Cdef} to 
show the third equality. 
One also checks the relation
\beq \label{def-H}
   H^{l  m} \equiv \int_{Y_4} \Psi^{l} \wedge * \bar \Psi^m  
   =  i \int_{Y_4} J \wedge \Psi^{l} \wedge \bar \Psi^m =  v^A d_A{}^{l  m} \ ,
\eeq
where we have used \eqref{starPsi} for the (1,2)-forms $\Psi^l$. This contraction gives precisely the 
positive definite metric of the complex scalars $N_l$ in \eqref{2Daction}.
It turns out to be convenient to write
\beq \label{splitH}
   H^{l  m}  = v^A d_A{}^{l  m} =   - \frac{1}{2} (\R f)^{ln} v^A C_{An}{}^m  \equiv - \frac{1}{2} (\R f)^{ln}\ \R\, h_{n}^m\ ,
\eeq 
where $h_{n}^m = - i t^A C_{An}{}^m$.  
Note that $H^{lm}$ thus depends non-trivially on the complex structure moduli $z^K$
through the holomorphic functions $f_{kl}$ and on the K\"ahler structure moduli $t^A$
through the holomorphic function $h_{n}^m $.
Second, we note that the modified derivative $DN_l$ appearing in \eqref{2Daction} is a shorthand for
\beq
  D N_l = d N_l - 2 \text{Re} N_m (\text{Re} f)^{mn} \partial_{z^K} (\text{Re} f_{nl} ) dz^K\ .
\eeq
Using this expression one easily reads off the coefficient function in front of $dN_l \wedge * dz^K$
and checks that it can be obtained by taking derivatives of a real function. In the next subsection 
we show that this is true for all terms in \eqref{2Daction} and discuss the connection with two-dimensional 
supersymmetry. 

\subsection{Comments on two-dimensional $\cN=(2,2)$ supergravity} \label{22dilatonSugra}

Having performed the dimensional reduction we next want to comment 
on the supersymmetry properties of  the 
action \eqref{2Daction}. As pointed out already 
in the previous subsection the  counting of covariantly constant spinors on 
the Calabi-Yau fourfold suggests that the two-dimensional effective theory admits 
$\cN=(2,2)$ supersymmetry. It was pointed out in \cite{Gates:2000fj} 
that, at least in the case of $h^{2,1}(Y_4)=0$ one expects to be 
able to bring the action \eqref{2Daction} into the standard form of 
an two-dimensional $\cN=(2,2)$ dilaton supergravity. In 
this work the dilaton supergravity action was constructed using 
superspace  techniques. Earlier works in this direction include \cite{Gates:1994gx,deWit:1992xr,Grisaru:1995dr}.
In the following we comment on this matching for $h^{2,1}(Y_4)=0$ 
and then discuss the general case in which $h^{2,1}(Y_4)>0$.

In order to display the supergravity actions we first have to introduce two sets 
of multiplets containing scalars in two-dimensions: (1) a set of chiral multiplets
with complex scalars $\phi^\kappa$, 
(2) a set of twisted-chiral multiplets with complex scalar $\sigma^A$.
In a superspace description these multiplets obey the two
inequivalent linear spinor derivative constraints leading to 
irreducible representations.   

To discuss the actions we first focus on the 
case $h^{2,1}(Y_4)=0$ and follow the constructions of \cite{Gates:2000fj}. 
For simplicity we will not include gaugings or a scalar potential. 
The superspace action used in \cite{Gates:2000fj} is 
given by 
\beq \label{suspace1}
  S^{(2)}_{\rm dil}  = \int d^2 x d^4 \theta E^{-1} e^{-2V - \cK}\ . 
\eeq
Here $E^{-1}$ is the superspace measure, $V$ is a real superfield with $V| = \varphi$ as lowest component,
and $\cK$ is a function of the chiral and twisted-chiral multiplets with lowest components $\phi^\kappa$ 
and $\sigma^A$, respectively. To display the bosonic part of the action \eqref{suspace1} 
we first set 
\beq
    e^{-2 \tilde \varphi} = e^{-2 \varphi - \cK}\ ,
\eeq
where $\cK (\phi^\kappa,\bar \phi^\kappa,\sigma^A,\bar \sigma^A)$ is evaluated as a function of 
the bosonic scalars. With this definition at hand one finds the bosonic action
\begin{align} \label{bosonic_dilatonsugra}
 S^{(2)}_{\rm dil} = \int e^{-2\tilde{\varphi}}  & \left( \frac{1}{2}R *1 + 2d \tilde{\varphi} \wedge * d\tilde{\varphi} 
 -  \cK_{\phi^\kappa \bar \phi^\lambda} \, d\phi^\kappa \wedge * d \bar {\phi}^\lambda
	  +  \cK_{ \sigma^A \bar  \sigma^B }\, d \sigma^A \wedge * d \bar {\sigma}^B \right.\nonumber \\
   & \ \ - \left. \cK_{\phi^\kappa \bar  \sigma^B} \, d \phi^\kappa \wedge d \bar \sigma^B
	  - \cK_{\sigma^A \bar \phi^\lambda} \, d \bar {\phi}^\lambda \wedge d\sigma^A \right)\ ,
\end{align}
where $\cK_{\phi^\kappa \bar \phi^\lambda } = \partial_{\phi^\kappa} \partial_{\bar \phi^\lambda} \cK $, 
$\cK_{ \phi^\kappa \bar \sigma^A } = \partial_{\phi^\kappa} \partial_{\bar \sigma^A} \cK $ 
with a similar notation for the other coefficients.
It is now straightforward to compare \eqref{bosonic_dilatonsugra} with the action 
\eqref{2Daction} for the case $h^{2,1}(Y_4)=0$, i.e.~in the absence of any complex scalars $N_l$. 
One first identifies 
\beq
  \tilde \varphi = \phi_{\rm IIA} \ ,\qquad \phi^K = z^K\ , \qquad \sigma^A = t^A\ ,
\eeq
and then infers that 
\beq \label{hatKsimple}
 \cK = - \log \int_{Y_4} \Omega \wedge \bar \Omega + \log \cV \ .
\eeq
Note that we find here a positive sign in front of the logarithm of $\cV$. This is related 
to the fact that there is an extra minus sign in the kinetic terms of the twisted-chiral 
fields $\sigma^A$ in \eqref{bosonic_dilatonsugra}. Clearly, the kinetic terms of the 
complex structure deformations $z^K$ and complexified 
K\"ahler structure deformations $t^A$ in the action \eqref{2Daction} have both positive definite kinetic 
terms.\footnote{Our discussion differs here from the one in \cite{Gates:2000fj}, where the 
sign in front of $\log \cV$ was claimed to be negative.}

Let us now include the complex scalars $N_l$. It is important to note that 
the action \eqref{2Daction} cannot be brought into the form \eqref{bosonic_dilatonsugra}.
In fact, we see in \eqref{2Daction} that the terms independent of the two-dimensional 
metric do not contain an $\phi_{\rm IIA}$-dependent pre-factor, while the terms of this type 
in \eqref{bosonic_dilatonsugra} do admit an $\tilde \varphi$-dependence. Any field
redefinition in \eqref{2Daction} involving the dilaton seems to introduce new 
undesired mixed terms that cannot be matched with \eqref{bosonic_dilatonsugra} either.
However, we note that the action \eqref{2Daction} actually can be 
brought to the form
\begin{align} \label{bosonic_dilatonsugra_extended}
 S^{(2)} = \int e^{-2\tilde{\varphi}} & \left( \frac{1}{2}R *1 + 2d \tilde{\varphi} \wedge * d\tilde{\varphi} 
 -  \tilde K_{\phi^\kappa \bar \phi^\lambda} \, d\phi^\kappa \wedge * d \bar {\phi}^\lambda
	  +  \tilde  K_{ \sigma^A \bar \sigma^B}\, d \sigma^A \wedge * d \bar {\sigma}^B \right.\nonumber \\
   & - \left. \tilde K_{ \phi^\kappa \bar  \sigma^B} \, d \phi^\kappa \wedge d \bar \sigma^B
	  - \tilde K_{ \sigma^A \bar \phi^\lambda} \, d \bar {\phi}^\kappa \wedge d\sigma^A \right)\ ,
\end{align}
where $\tilde K$ is now allowed to be dependent on $\tilde \varphi$ and given by 
\beq \label{tildeK}
    \tilde K = \cK + e^{2 \tilde \varphi} \cS\ , 
\eeq
Similar to $\cK$, the new function $\cS$ is allowed to depend on the chiral and twisted-chiral scalars $\phi^\kappa, \sigma^A$,
but is taken to be independent of $\tilde \varphi$. The action \eqref{bosonic_dilatonsugra_extended}
trivially reduces to \eqref{bosonic_dilatonsugra} for $\cS=0$. Note that the new terms 
induced by $\cS$ do not scale with $e^{-2\tilde \varphi}$.
Comparison with \eqref{2Daction} reveals that one can identify
\beq \label{id_fields}
  \tilde \varphi = \phi_{\rm IIA} \ ,\qquad \phi^\kappa = (z^K,N_l)\ , \qquad \sigma^A = t^A\ ,
\eeq
and introduce the generating functions 
\begin{align} \label{idhatKS}
   &\cK = - \log \int_{Y_4} \Omega \wedge \bar \Omega + \log \cV \ , \\
   & \cS =     H^{l   k} \ \R\, N_l\ \R N_k\  ,\qquad  H^{l   k} \equiv v^A d_A{}^{l   k}  \ . \nn
\end{align}
To show this, it is useful to note that $d_A{}^{l  k}$ can be evaluated 
as in \eqref{evaluate_d} and depends on the complex structure moduli through 
the holomorphic function $f_{mn}(z)$ only. 

Let us close this subsection with two remarks. First, note 
that \eqref{bosonic_dilatonsugra_extended} is
expected to be compatible with $\cN=(2,2)$ supersymmetry and gives an extension 
of the two-dimensional dilaton supergravity action \eqref{suspace1}. 
A suggestive form of the extended superspace action is  
\beq \label{suspace2}
  S^{(2)} = \int d^2 x d^4 \theta E^{-1} \big(e^{-2V - \cK} + \cS\big)\ ,
\eeq
where $\cS$ is now evaluated as a function of the chiral and twisted-chiral superfields.
It would be interesting to check that \eqref{suspace2} indeed correctly reproduces the 
bosonic action \eqref{bosonic_dilatonsugra_extended} with $\tilde K$ as in \eqref{tildeK}. 

Second, the action \eqref{bosonic_dilatonsugra_extended} with the identification \eqref{id_fields}
can also be straightforwardly obtained by dimensionally reducing M-theory, or rather 
eleven-dimensional supergravity, first on $Y_4$ and then on an extra circle of radius $r$. 
The reduction of M-theory on $Y_4$ was carried out in \cite{Haack:1999zv,Haack:2001jz}. We give the resulting 
three-dimensional action in \eqref{3daction} and briefly recall 
this reduction in \autoref{Mreduction_details} when considering applications to F-theory.
Using the standard relation of eleven-dimensional supergravity on a circle and Type IIA supergravity,
one straightforwardly identifies 
\beq \label{id_vA}
   r = e^{-2\phi_{\rm IIA}}\ , \qquad     e^{2 \phi_{\rm IIA}} v^A = \frac{v^A_{\rm M}}{\mathcal{V}_{\rm M}} \equiv L^A\ , 
\eeq
where $v^A_{\rm M}$ and $\cV_{\rm M}$ are the analogs of $v^A$ and $\cV$
used in the M-theory reduction. Note that the scalars $L^A$ are the appropriate fields to appear in three-dimensional 
vector multiplets. Inserting the identification \eqref{id_vA} 
into \eqref{tildeK} together with \eqref{id_fields}, \eqref{idhatKS} one finds
\beq  \label{def-tildeKM}
   \tilde K^{\rm M} =  - \log \int_{Y_4} \Omega \wedge \bar \Omega + \log \Big(\frac{1}{4!} \cK_{ABCD}L^A L^B L^C L^D \Big) +    L^A d_A{}^{l  k}\  \R N_l\  \R N_k\ ,
\eeq
where we have dropped the logarithm containing the circle radius. Indeed $\tilde K^{\rm M}$ agrees 
precisely with the result found in \cite{Haack:1999zv,Haack:2001jz,Grimm:2010ks} from the M-theory reduction. 
The general discussion of the circle reduction of a three-dimensional un-gauged $\cN=2$ supergravity theory 
to a two-dimensional $\cN=(2,2)$ supergravity theory can be found in \autoref{3d-2dreduction}.

\subsection{Legendre transforms from chiral and twisted-chiral scalars} \label{sec_Legendre}

In this subsection we want to introduce an operation that allows to 
translate the dynamics of certain chiral multiplets to twisted-chiral multiplets and 
vice versa. More precisely, we will assume that some of the scalars, say the scalars 
$\lambda_l$, in the $\cN=(2,2)$ supergravity action have continuous shift 
symmetries, i.e.~$\lambda_l \rightarrow \lambda_l + c_l$ for constant $c_l$. 
These scalars therefore only appear with derivatives $d\lambda^l$ in the action. 
By the standard duality of massless $p$-forms to $(D-p-2)$-forms in $D$ dimensions, 
one can then replace the scalars $\lambda_l$ by 
dual scalars $ \lambda'^{\, l}$. Accordingly, one has to adjust 
the complex structure on the scalar field space by performing a 
Legendre transform. 
In the following we will give representative  examples
of how this works in detail. We will see that this duality, in particular 
as described in the first example, becomes crucial 
in the discussion of mirror symmetry of \autoref{mirror_section}.

As a first example, let us consider the above theory with complex scalars 
$z^K,N_l$ in chiral multiplets and complex scalars $t^A$ in twisted-chiral 
multiplets. The kinetic potential for these fields $\tilde K$
was given in \eqref{tildeK} with \eqref{idhatKS}.
Two facts about this example are crucial for the following discussion. 
First, the fields $N_l$ admit a shift symmetry $N_l \rightarrow N_l + i c_l$ in the 
action, i.e.~the kinetic potential $\tilde K$ given in \eqref{idhatKS} is independent of $N_l - \bar N_l$.
Second, the $N_l$ only appear in the term $\cS$ of 
the kinetic potential and thus carry no dilaton pre-factor in the 
action. One can thus straightforwardly 
dualize $N_l - \bar N_l$ into real scalars $\lambda'^{\, l}$. 
The new complex scalars $N'^{\, l}$ are then given by  
\beq \label{N'}
   N'^{\, l} =  \frac{1}{2}\frac{\partial \cS}{\partial \R \, N_l}+ i \lambda'^{\, l}\ ,
   \eeq
where we have included a factor of $1/2$ for later convenience. 
Furthermore, the new kinetic potential $\tilde K'$ is now a function 
of $z^K,\ t^A,\ N'^{\, l}$ and given by the Legendre transform
\beq \label{tildeK'}
   \tilde K'  = \tilde K - 2\, e^{2\tilde \varphi} \R \, N'^{\, l}  \R \, N_l \ ,
\eeq
where $ \R \, N_l $ has to be evaluated as a function of $\R \, N'^{\, l}$
and the other complex fields by solving \eqref{N'} for $\R \, N_l$.
One now checks that the scalars $N'^{\, l}$ actually reside in 
twisted-chiral multiplets. Using the  transformation \eqref{N'} and 
\eqref{tildeK'} in the action \eqref{bosonic_dilatonsugra_extended} simply yields a dual description in which 
certain chiral multiplets are consistently replaced by twisted-chiral multiplets.  
It is simple to  evaluate \eqref{N'}, \eqref{tildeK'} for $\cS$ given in
\eqref{idhatKS} to find
\bea
  &N'^{\, l} &=  H^{lm}\ \R\, N_m  + i \lambda'^{\, l}\ , \label{N'ex} \\
  &\tilde K'  &= \cK -   e^{2 \phi_{\rm IIA}} H_{kl} \ \R\, N'^{\, k} \ \R\, N'^{\, l}\ ,\label{tildeK'ex}
\eea
where $H^{lm}$ is the inverse of the matrix $H_{lm}$ introduced in \eqref{def-H}, \eqref{splitH}.
It is interesting to realize that upon inserting \eqref{N'ex} into \eqref{tildeK'ex} 
one finds that $\tilde K'$ evaluated as a function 
of $N_k$ only differs by a minus sign in front of the term linear in $e^{2 \phi_{\rm IIA}}$
from the original $\tilde K$. This simple transformation arises from the 
fact that $\tilde K$ is only quadratic in the $N_k$. This observation 
will be crucial again in the discussion of mirror symmetry in \autoref{mirror_section}.

As a second example, we briefly want to discuss a dualization that 
transforms all multiplets containing scalars to become chiral. 
The detailed computation for a general $\cN=(2,2)$ setting can be 
found in \autoref{detailed_dual}. 
For the example of \autoref{22dilatonSugra} we focus on the twisted-chiral 
multiplets with complex scalars 
$t^A$. These admit a shift symmetry $t^A \rightarrow t^A + c^A$ for 
constant $c^A$, such that $\R\, t^A$ only appears with derivatives in 
the action. Accordingly, the kinetic potential $\tilde K$
is independent of $t^A + \bar t^A$ as seen in  \eqref{tildeK} with \eqref{idhatKS}. 
Due to the shift symmetry we can dualize the scalars $t^A + \bar t^A$
to scalars $\rho_A$. However, note that by using the kinetic potential  \eqref{tildeK}, \eqref{idhatKS}
there are couplings of $t^A$ 
in \eqref{bosonic_dilatonsugra_extended} that have a dilaton 
factor $e^{\tilde \varphi}$, and others that are independent 
of $e^{\tilde \varphi}$. 
This seemingly prevents us from performing a straightforward Legendre transform 
to bring the resulting action to the form \eqref{bosonic_dilatonsugra_extended}
with only chiral multiplets. Remarkably, the special properties of the 
kinetic potential \eqref{tildeK}, \eqref{idhatKS}, however, allow us to nevertheless achieve this goal
as we will see in the following. 

The action \eqref{bosonic_dilatonsugra_extended} for a setting with only chiral multiplets with 
complex scalars $M^I$ takes the form 
\beq \label{S2Kaehler}
   S^{(2)} = \int e^{-2\tilde{\varphi}} \left( \frac{1}{2}R *1 + 2d \tilde{\varphi} \wedge * d\tilde{\varphi} 
 -  \mathbf{K}_{M^I \bar M^J} \, dM^I\wedge * d \bar M^J \right)\ ,
\eeq
where $  \mathbf{K}_{M^I \bar M^J} = \partial_{M^I} \partial_{\bar M^J} \mathbf{K}$.
In other words, the potential $\mathbf{K}$ is in this case actually a K\"ahler potential 
on the field space spanned by the complex coordinates $M^I$. 
For our example  \eqref{tildeK}, \eqref{idhatKS} the scalars $M^I$ 
consist of $z^K$, $N_l$, and $T_A$, where $T_A$ are the duals of the 
complex fields $t^A$.
We make the following Ansatz for the dual coordinates 
$T_A$
\beq \label{TA_Ansatz1}
   T_A =  e^{-2\tilde \varphi} \frac{\partial \tilde K}{\partial \I\, t^A} + i \rho_A=
    e^{-2\tilde \varphi} \frac{\partial \cK }{\partial \I\, t^A}+ \frac{\partial \cS}{\partial \I\, t^A} + i \rho_A \ , 
\eeq
  and the dual potential $\mathbf{K}$ 
\beq \label{TA_Ansatz2}
    \mathbf{K} = \tilde K - e^{2\tilde \varphi} \R \, T_A \I \, t^A\ .  
\eeq
These expressions describe the standard Legendre transform for $\I \,t^A$, but 
crucially contain dilaton factors $e^{2\tilde \varphi}$. This latter fact 
allows to factor out $e^{-2\tilde \varphi}$ as required in \eqref{S2Kaehler},
but requires to perform a two-dimensional Weyl rescaling as we will discuss 
below. Using  \eqref{tildeK} with \eqref{idhatKS} one 
straightforwardly evaluates
\bea \label{TAbfK}
   &T_A &= e^{-2\phi_{\rm IIA}} \frac{1}{3!} \frac{\cK_{A}}{\cV} + d_{A}{}^{kl}\ \R\, N_l \ \R\, N_k  +  i \rho_A \ , \\
   & \mathbf{K} &= - \log \int_{Y_4} \Omega \wedge \bar \Omega + \log \cV \ .
\eea
Clearly, upon using the map \eqref{id_vA} 
this result is familiar from the study of M-theory compactifications 
on Calabi-Yau fourfolds \cite{Haack:1999zv,Haack:2001jz,Grimm:2010ks}. Also note that the contribution 
$\cS$ present in the kinetic potential \eqref{tildeK} is removed by the Legendre transform
in $\mathbf{K} $  and reappears in a more involved definition of the coordinates $T_A$. 

At first it appears that \eqref{TA_Ansatz1} induces new mixed terms involving 
one $d\tilde \varphi$ due to the dilaton dependence in front of the derivatives of $\cK$. 
Interestingly, these can be removed by a two-dimensional Weyl rescaling 
if $\cK$ satisfies the conditions 
\beq \label{hatKcond}
  \cK_{t^A} \, \cK^{t^A \bar t^B}\, \cK_{\bar t^B} = k\ ,
  \qquad  \cK_{v^A} \ d \I t^A = d f \ , 
\eeq
for some constant $k$ and some real field dependent function $f$. 
In this expression $\cK^{t^A \bar t^B}$ is the inverse of $\cK_{t^A \bar t^B}$ 
and $\cK_{v^A} \equiv \partial_{\I\, t^A} \cK$.
In fact, one can perform the rescaling $\tilde g_{\mu \nu} = e^{2 \omega} g_{\mu \nu}$,
which transforms the Einstein-Hilbert action as 
\beq \label{Weyl-EH}
   \int  e^{-2 \tilde \varphi} \frac{1}{2} \tilde R \ \tilde *1  = 
    \int e^{-2 \tilde \varphi} \left( \frac{1}{2} R \  *1 - 2 d\omega \wedge * d\tilde \varphi  \right) \ ,
\eeq
while leaving all other terms invariant. 
Using \eqref{Weyl-EH} to absorb the mixed terms one needs to chose
\beq
 \omega = - \frac{k}{2} \tilde \varphi - \frac{f}{2} \ .
\eeq
The details of this computation can be found in \autoref{detailed_dual}. 
Indeed, for the example \eqref{idhatKS} one finds $f= \text{log}\ \cV$
and $k=-4$. Remarkably, the condition \eqref{hatKcond}
essentially states that $\cK$ has to satisfy a no-scale like condition. 
A recent discussion and further references on the subject of 
studying four-dimensional supergravities satisfying 
such conditions can be found in \cite{Ciupke:2015ora}.

\section{Mirror symmetry at large volume/large complex structure} \label{mirror_section}

In \autoref{IIAreduction} we have determined the two-dimensional action obtained from 
Type IIA supergravity compactified on a Calabi-Yau fourfold. We commented 
on its $\cN=(2,2)$ supersymmetry structure which relies on the proper identification 
of chiral and twisted-chiral multiplets in two dimensions. In this section 
we are exploring the action of mirror symmetry. More precisely, we consider 
pairs of geometries $Y_4$ and $\hat Y_4$ that are mirror manifolds \cite{Greene:1993vm,Mayr:1996sh,Klemm:1996ts}.
From a string theory world-sheet perspective one expects the two theories 
obtained from string theory on $Y_4$ and $\hat Y_4$ to be dual. This implies 
that after finding the appropriate identification of coordinates the two-dimensional effective 
theories should be identical when considered at dual points in moduli space. 
We will make this more precise for the large volume and large complex structure 
point in this section. Note that in contrast to mirror symmetry for Calabi-Yau threefolds the mirror 
theories encountered here are both arising in Type IIA string theory.\footnote{This can be seen 
immediately when employing the SYZ-understanding of mirror symmetry as T-duality \cite{Strominger:1996it}. 
Mirror symmetry is thereby understood as T-duality along half of the compactified 
dimensions, i.e.~$Y_4$ is argued to contain real four-dimensional tori along which T-duality can be 
performed. Clearly, this inverts an even number of dimensions for Calabi-Yau fourfolds.}

\subsection{Mirror symmetry for complex and K\"ahler structure}

Mirror symmetry arises from the observation that the conformal field theories associated with
$Y_4$ and $\hat Y_4$ are equivalent. 
It describes the identification of  
Calabi-Yau fourfolds $Y_4$, $\hat Y_4$
with Hodge numbers
\beq
     h^{p,q}(Y_4) = h^{4-p,q}(\hat {Y}_4)\ .
\eeq
Note that this particularly includes the non-trivial conditions 
\bea 
   &h^{1,1}(Y_4) = h^{3,1} (\hat Y_4) \ , \qquad h^{3,1}(Y_4) = h^{1,1} (\hat Y_4)\ ,&  \label{h11=h31} \\
    &h^{2,1}(Y_4) = h^{2,1}(\hat Y_4)\ .  \label{h21=h21} &
\eea
The first identification \eqref{h11=h31} together with the observations made in 
\autoref{IIAreduction} implies that mirror symmetry exchanges K\"ahler structure deformations 
of $Y_4$ ($\hat Y_4$) with complex structure deformations of $\hat Y_4$ ($Y_4$). 
 Accordingly one needs to 
exchange chiral multiplets and twisted-chiral multiplets in the effective $\cN=(2,2)$ supergravity 
theory. The second identification \eqref{h21=h21} seems to suggest that for the fields $N_l$
the mirror map is trivial. However, as we will see in \autoref{3form_mirror} this is not the case 
and one has to equally change from a chiral to a twisted-chiral description.  

To present a more in-depth discussion of mirror symmetry we first 
need to introduce some notation. All fields and couplings obtained  
by compactification on $Y_4$ are denoted as in \autoref{IIAreduction}. 
To destinguish them from the quantities obtained in the  $\hat Y_4$ 
reduction we will dress the latter with a hat. In particular for the fields we write
\bea
   Y_4 :& \quad & \phi_{\rm IIA}\, ,\ t^A\, ,\ z^K \,  ,\ N_l \ , \\
   \hat Y_4 :& \quad & \hat \phi_{\rm IIA}\, ,\ \hat t^K\, ,\ \hat z^A \,  ,\ \hat N_l \ . \nn
\eea 
Note that we have exchanged the indices on $\hat t^K$ and $\hat z^A$
in accordance with the fact that complex structure and K\"ahler structure 
deformations are interchanged by mirror symmetry.  In other words,
$K= 1, \ldots, h^{1,1}(\hat Y_4)$ and $A = 1, \ldots, h^{3,1}(\hat Y_4)$
is compatible with the previous notation due to \eqref{h11=h31}.
Similarly we will adjust the notation for the couplings. For example, 
the functions introduced in \eqref{idhatKS} and \eqref{def-f1}, \eqref{def-Psil} are 
\bea
   Y_4 :& \quad &  f_{mn}(z) \, ,\ H^{mn}(v,z) \, , \\
   \hat Y_4 :& \quad &\hat f_{mn}(\hat z) \, , \ \hat H^{mn}(\hat v,\hat z)\,  .
\eea
The functional form of the various couplings will in general differ for 
 $Y_4$ and $\hat Y_4$. A match of the two mirror-symmetric 
effective theories should, however, be possible when identifying 
the mirror map, which we denote formally by $\cM[\cdot]$.

We want to focus on the sector of the theory independent of the three-forms. 
Recall that in the two-dimensional effective theory obtained from 
$Y_4$ the kinetic terms of the complex structure moduli $z^K$ and 
K\"ahler structure moduli $t^A$ are obtained from the kinetic potential \eqref{hatKsimple}, \eqref{idhatKS} as
\begin{align} \label{hatKY4}
 \cK(Y_4) =  \log \Big( \frac1{4!} \cK_{ABCD}\, \I \,t^A \, \I\, t^B \, \I \, t^C\, \I \, t^D \Big)- \log \int_{Y_4} \Omega \wedge \overline{\Omega}
\end{align}
when used in the action \eqref{bosonic_dilatonsugra}. 
Mirror symmetry exchanges the K\"ahler moduli $t^K$ of $Y_4$ with the 
complex structure moduli $\hat z^K$ of $\hat Y_4$. The expression
\eqref{hatKY4} was computed at the large volume 
point in K\"ahler moduli space, i.e.~with the assumption that $\I \, t^A \gg 1$ in string units. 
Accordingly one has to evaluate  $\cK(\hat Y_4)$ at the large complex structure 
point as 
\beq
   \int_{\hat Y_4} \hat \Omega \wedge \overline{ \hat \Omega} = 
   \frac1{4!} \cK_{ABCD}\, \I \,\hat z^A \, \I \,\hat z^B \, \I \,\hat z^C\, \I \,\hat z^D\ ,
\eeq
where now $\I \, \hat z^A \gg 1$.
Similarly, one has to proceed for the K\"ahler moduli part of the kinetic potential $\cK(\hat Y_4)$
and evaluate $\cK(Y_4)$ at the large complex structure point 
\beq
   \int_{Y_4}   \Omega \wedge \overline{   \Omega} = \frac1{4!} \hat \cK_{KLMN}\, \I\,   z^K \, \I  \, z^L \, \I  \, z^M\, \I  \, z^N\ ,
\eeq
where $\hat \cK_{KLMN}$ are now the quadruple intersection numbers on the geometry $\hat Y_4$. 
Therefore, at the large volume and large complex structure point the two effective theories obtained 
from $Y_4$ and $\hat Y_4$ are identified under the mirror map
\beq \label{mirror_tz}
 \mathcal{M} \big[t^A \big]= \hat{z}^A\ , \quad \mathcal{M}\big[z^K\big]= \hat{t}^K\ ,
\eeq
and  
\beq \label{mirror_phiIIA}
 \mathcal{M} \big[ \cK(Y_4) \big] = - \cK(\hat Y_4)\ , \qquad \cM \big[ \phi_{\rm IIA}\big] = \hat \phi_{\rm IIA} \ .
\eeq
It is important to stress that a sign change occurs when applying the mirror map to $\cK$. 
This can be traced back to the fact that scalars in chiral and twisted-chiral multiplets have different 
sign kinetic terms in the actions \eqref{bosonic_dilatonsugra}, \eqref{bosonic_dilatonsugra_extended}. 
The quantum corrections to $\cK$ were discussed using mirror symmetry in \cite{Greene:1993vm,Mayr:1996sh,Klemm:1996ts,Grimm:2009ef} 
and localization in \cite{Honma:2013hma,Halverson:2013qca} (using and extending
the results of \cite{Benini:2012ui,Doroud:2012xw,Jockers:2012dk}). 

\subsection{Mirror symmetry for non-trivial three-forms} \label{3form_mirror}

Let us next include the moduli $N_l$ arising for Calabi-Yau fourfolds $Y_4$ with non-vanishing 
$h^{2,1}(Y_4)$. In \autoref{IIAreduction} we have seen that these complex scalars  
are part of chiral multiplets. Their dynamics was described by 
the real function $\cS$ in the kinetic potential $\tilde K$ given in \eqref{tildeK} and \eqref{idhatKS}. 
For completeness we recall that
\beq \label{SY4}
  \cS(Y_4) =   H^{l   k}\ \R\, N_l\ \R\,N_k\ ,\qquad  H^{l   k} \equiv v^A d_A{}^{l   k} \ , 
\eeq
where $d_A{}^{l   k} $ is a function of the complex structure moduli of $Y_4$. 
Mirror symmetry should map the fields $N_l$ to scalars $\hat N_l$ arising 
in the reduction on the mirror Calabi-Yau fourfold $\hat Y_4$, i.e.~one 
should have 
\beq \label{cMN}
   \cM \big[ N_l \big] = Q_{l}( \hat N, \hat z, \hat t)\ ,
\eeq
where we have allowed the image of $N_l$ to be a non-trivial 
function that will be determined in the following. 
In fact, note that the map cannot be as simple as $\cM(N_l) = \hat N_l$.
As already pointed out in  \cite{Gates:2000fj} 
the mirror duals $\cM(N_l)$ need to be, in contrast to the $N_l$, parts 
of \textit{twisted}-chiral multiplets. To 
achieve this we need to  use the 
results of \autoref{sec_Legendre}.

Let us therefore consider the reduction on $\hat Y_4$ using the same notation as 
in \autoref{IIAreduction} but with hatted symbols. The two-dimensional theory 
will contain a set of complex scalars $\hat N_l$ that reside in chiral multiplets. 
We can transform them to scalars in twisted-chiral multiplets using \eqref{N'ex} 
and \eqref{tildeK'ex}. In other words, we find a dual description with 
scalars $\hat N'^l$ defined as 
\beq \label{hatN'}
   \hat N'^{\, l} =  \hat H^{lm}\, \R\, \hat N_m  + i \hat \lambda'^{\, l}\ , 
\eeq
where $ \hat H^{lm}$ is a function of the mirror complex structure moduli $\hat z^A$
and K\"ahler moduli $\hat v^K$. The dual kinetic potential takes the form
\beq \label{tildeK'hatY4}
  \tilde K'(\hat Y_4) =  \cK(\hat Y_4) -  e^{2 \hat \phi_{\rm IIA}} \hat H_{kl}\ \R\, \hat N'^{\, k} \ \R\, \hat N'^{\, l}\ . 
\eeq
The mirror map \eqref{mirror_tz}, \eqref{mirror_phiIIA} and  \eqref{cMN} 
exchanges chiral and twisted-chiral states and therefore has to take the form 
\bea
  & \cM \big[ N_l \big]& = \hat N'^{\, l}(\hat N,\hat z,\hat t)\ , \qquad \mathcal{M} \big[ t^A\big]= \hat{z}^A\ , 
  \quad \mathcal{M}\big[ z^K \big] = \hat{t}^K\ ,  \label{full_mirror1} \\
  & \mathcal{M}\big[ \tilde K(Y_4)\big] &= - \tilde {K}'(\hat Y_4)\ , \qquad \cM\big[ \phi_{\rm IIA}\big] = \hat \phi_{\rm IIA}\ . 
  \label{full_mirror2}
\eea
and is evaluated as a function of $\hat N_l$, $\hat z^A$ and $\hat t^K$ by using \eqref{hatN'}.

Using these insights we are now able to infer the mirror image of the function $H_{mn}$ appearing 
in $\tilde K(Y_4)$. 
To do that, we apply the mirror map to the kinetic potential $\tilde K$. Note 
that 
\beq
   \cM \big[ \tilde K(Y_4) \big] = - \cK(\hat Y_4) + e^{2 \hat \phi_{\rm IIA}} \cM \big[ \cS(Y_4) \big]\ , 
\eeq
where we have used \eqref{mirror_phiIIA}. Furthermore, we insert \eqref{full_mirror2} into \eqref{SY4}
to find 
\beq
 \cM \big[ \cS(Y_4) \big] = \sum_{k,l} \cM \big[ H^{kl} \big] \R\, \hat N'^k  \, \R\, \hat N'^l\  .
\eeq
We next apply \eqref{full_mirror2} together with \eqref{tildeK'hatY4} which requires 
\beq
     \sum_{k,l} \cM \big[ H^{kl}\big] \R\, \hat N'^k  \, \R\, \hat N'^l \overset{!}{=} \hat H_{kl}\ \R\, \hat N'^{\, k} \ \R\, \hat N'^{\, l}\ ,
\eeq 
and thus enforces
\beq \label{cMH}
    \cM \big[ H^{kl} \big]  \overset{!}{=} \hat H_{kl}\ .
\eeq
We therefore find that the mirror map actually identifies $H^{kl}$ with the \textit{inverse} $\hat H_{kl}$ of $\hat H^{kl}$. 
This inversion is crucial and stems from the exchange of chiral an twisted-chiral multiplets under mirror symmetry. 
In the final part of this section we evaluate the condition \eqref{cMH} at the large complex structure 
point, since $H^{kl}$ given in \eqref{SY4} was computed  at large volume.

Using the mirror map we are now able to determine the holomorphic function $f_{kl}$ appearing 
in the definition of $H_{kl}$ at the large complex structure point.
Note that \eqref{splitH} translates on $Y_4$ and $\hat Y_4$ to 
\bea \label{recall_HhatH}
   &H^{lm} &=  - \frac{1}{2}  (\R f)^{ln} \ \R\, h_n^m \ , \qquad   h_n^m = - i t^A C_{An}{}^m \ , \qquad \\
   &\hat H^{lm} &=  -\frac{1}{2}  (\R \hat f)^{ln} \ \R\, \hat h_n^m \ ,\qquad   \hat h_n^m =-i\hat t^K \hat C_{Kn}{}^m\ , \nn
\eea
where on the mirror geometry we introduced the intersection numbers
\begin{equation}
  \hat C_{Kn}{}^m = \int_{\hat Y_4} \hat{\omega}_K \wedge \hat{\alpha}_n \wedge \hat{\beta}^m\ .
\end{equation}
Using \eqref{full_mirror1}, \eqref{full_mirror2}, \eqref{cMH}, and \eqref{recall_HhatH} in the
mirror map one infers that a possible identification is \footnote{Note that 
in general the basis $(\alpha_l,\beta^k)$ might not directly map to $(\hat \alpha_l,\hat \beta^k)$
on the mirror geometry $\hat Y_4$. In this expression we have assumed that there is no non-trivial 
base change under mirror symmetry. }
\beq
   \R f_{nm}= \I z^K \hat C_{Kn}{}^m\ .
\eeq
By holomorphicity of $f_{nm}$ we finally conclude
\beq   \label{lin_fresults}
f_{nm}= -i z^K \hat C_{Kn}{}^m
\eeq
Having determined the function $f_{mn}$ at the large complex 
structure point we have established a complete match of the 
two two-dimensional effective theories obtained from $Y_4$ and
$\hat Y_4$ under the mirror map $\cM[\cdot]$. The result \eqref{lin_fresults}
is not unexpected. In fact, from the variation of Hodge-structures one 
could have expected a leading linear dependence on $z^K$. Furthermore, 
we will find agreement with a dual Calabi-Yau threefold result when using 
the geometry $Y_4$ as F-theory background and performing the orientifold limit. 
This will be the task of the final section of this work.

\section{Applications for F-theory and Type IIB orientifolds} \label{F-theoryapp}

In this section we want to apply the result obtained by using 
mirror symmetry to compactifications of F-theory and their orientifold limit.  
The F-theory effective action is studied via the M-theory to 
F-theory limit. Therefore, we will briefly review in \autoref{Mreduction_details} 
the dimensional reduction of M-theory on a smooth Calabi-Yau fourfold including three-form moduli. 
This  reduction was already performed in \cite{Haack:2001jz}, but we will use the insights we have gained in the previous sections to include the three-form moduli more conveniently. 
In \autoref{MFlimit} we will then restrict to a certain class 
of elliptically fibered Calabi-Yau fourfolds and perform the M-theory to F-theory limit. 
This allows us to identify the characteristic data determining the
 four-dimensional $ \cN = 1 $ F-theory effective action in terms of the geometric 
 quantities of the internal space \cite{Grimm:2010ks}. We note that for certain fourfolds the holomorphic 
 function $f_{kl}$ lifts to a four-dimensional gauge coupling function. 
 Starting from these F-theroy settings we will then perform the weak string coupling limit in \autoref{orientifold_limit}.
 In this limit $f_{kl}$ can be partially computed by using mirror symmetry for Calabi-Yau threefolds
 and we show compatibility with the fourfold result of \autoref{mirror_section}.

\subsection{M-theory on Calabi-Yau fourfolds} \label{Mreduction_details}

In this subsection we review the dimensional reduction of M-theory on a Calabi-Yau fourfold $ Y_4 $ in the large volume limit without fluxes. The ansatz here is similar to the one used for Type IIA supergravity in \autoref{IIAreduction_details}.

We start with eleven-dimensional supergravity as the low-energy limit of M-theory. 
Its bosonic two-derivative action is given by
\begin{equation} \label{11Daction}
S^{(11)} = \int \frac{1}{2} \check R\ \check\ast 1 - \frac{1}{4} \check F_4 \wedge\check \ast \check F_4 - \frac{1}{12}\check C_3 \wedge \check F_4 \wedge \check F_4\ ,
\end{equation}
with $\check F_4 = d \check C_3 $ the eleven-dimensional three-form field strength. This will be dimensionally reduced on the background
\beq
   d\check s ^2 = \eta^{(3)}_{\mu \nu} dx^\mu dx^\nu + g^{(8)}_{mn} dy^m dy^n\ , 
\eeq
where $ \eta^{(3)} $ is the metric of three-dimensional Minkowski space-time $ \mathbb{M}_{2,1} $ and $ g^{(8)} $ the metric of the Calabi-Yau fourfold $ Y_4 $. This is the analog to \eqref{metric-ansatz}
and, as we briefly discussed at the end of \autoref{22dilatonSugra}, the Type IIA supergravity vacuum can be obtained by 
a circle-reduction of this Ansatz. 

To perform the dimensional reduction one inserts similar expansions of \eqref{Kaehlermoduli}, 
\eqref{CSmoduli} and \eqref{BC-expand} into the eleven-dimensional action \eqref{11Daction}. For the metric deformations consisting of K\"ahler and complex structure deformations, this is exactly the same as \eqref{Kaehlermoduli} and
\eqref{CSmoduli}, hence we obtain $ h^{1,1}(Y_4) $ real scalars $ v^A_{\rm M}$ by expanding the M-theory K\"ahler form $J_{\rm M}$ as
\beq \label{JM}
   J_{\rm M} = v^A _{\rm M}\omega_A
\eeq
and $ h^{3,1}(Y_4) $ complex scalars $ z^K $ in three dimensions. Since the eleven-dimensional three-form $\check C_3 $ is the common origin of the Type IIA fields 
$\check {B}_2,\ \check{C}_3$, we expand
\beq \label{11Dthree-form}
\check C_3 = V^A \wedge \omega_A + N_l \Psi^l + \bar{N}_l \bar{\Psi}^l \, .
\eeq
This yields  $ h^{2,1}(Y_4) $ three-dimensional complex scalars $ N_l $ and $ h^{1,1}(Y_4) $ vectors $ V^A $.
The latter combine with the real scalars $ v^A_{\rm M} $ into three-dimensional vector multiplets, whereas $ z^K, N_l $ give 
rise to three-dimensional  chiral multiplets. Combining the expansions \eqref{Kaehlermoduli}, 
\eqref{CSmoduli} and \eqref{11Dthree-form} with the action \eqref{11Daction} by using the notation of \autoref{IIAreduction_details} and \autoref{22dilatonSugra} we thus obtain the three-dimensional effective action 
\footnote{The action has been Weyl-rescaled to the three-dimensional Einstein 
frame by introducing ${g}^{\rm new}_{\mu \nu} = \mathcal{V}^{-2} g^{\rm old}_{\mu \nu} $}
\begin{align} \label{3daction}
S^{(3)} &= \int \frac{1}{2} R \ast 1 - G_{K\overline{L}} dz^K \wedge \ast d\overline{z}^L 
							- \frac{1}{2} d \log \mathcal{V}_{\rm M} \wedge \ast d \log \mathcal{V}_{\rm M} -  G_{A  B}^{\rm M} dv^A_{\rm M} \wedge \ast dv^B_{\rm M} \nonumber \\
					 &\quad - \frac{1}{2}v^A_{\rm M}\, d_{A}{}^{l  k} DN_l \wedge \ast D\overline{N}_k 
					 -  \mathcal{V}_{\rm M}^2  G^{\rm M}_{A B} dV^A \wedge \ast dV^B \nonumber \\
					 &\quad + \frac{i}{4}  d_A{} ^{l k} dV^A \wedge \big(N_l D\overline{N}_k  - \overline{N}_k DN_l \big) \ .
\end{align}
Note the $G_{AB}^{\rm M}$ takes the same functional form as \eqref{GABmetric}, but uses the 
M-theory K\"ahler structure deformations $v^A_{\rm M}$.

The three-dimensional action given in \eqref{3daction} is an $ \cN = 2 $ supergravity theory.
The proper scalars in the vector multiplets are denoted by $L^A$ and 
are expressed in terms of the $v^A_{\rm M}$ as $L^A = \frac{v^A_{\rm M}}{\cV_{\rm M}}$, 
as  already given in \eqref{id_vA}. The complex scalars in the chiral multiplets 
are collectively denoted by $\phi^\kappa = (z^K, N_l)$.
The action \eqref{3daction} can then be written using a kinetic potential $\tilde K^{\rm M}$ as
\begin{align} \label{chirallinear3D}
S^{(3)} = \int & \frac{1}{2} R^{(3)} \ast 1   + \frac{1}{4} \tilde{K}_{L^A L^B}^{\rm M} dL^A \wedge \ast dL^B 
     + \frac{1}{4} \tilde{K}^{\rm M}_{L^A L^B} dV^A \wedge \ast dV^B \nonumber \\
					 &  
					- \tilde{K}^{\rm M}_{\phi^\kappa \bar \phi^\lambda}\, d\phi^\kappa \wedge \ast d\bar  \phi^\lambda +  dV^A \wedge \I (\tilde K^{\rm M}_{L^A \phi^\kappa} d \phi^\kappa )\ , 
\end{align}   
where $\tilde{K}^{\rm M}_{L^A L^B}  = \partial_{L^A} \partial_{L^B} \tilde K$, 
$\tilde{K}^{\rm M}_{\phi^\kappa \bar \phi^\lambda} = \partial_{\phi^\kappa} \partial_{\bar \phi^\lambda} \tilde K^{\rm M}$, 
and $\tilde K^{\rm M}_{L^A \phi^\kappa} =  \partial_{L^A} \partial_{\phi^\kappa} \tilde K^{\rm M} $.
Comparing \eqref{3daction} with \eqref{chirallinear3D} the 
kinetic potential obtained for this M-theory reduction therefore reads 
\beq \label{def-tildeKM_Recall}
     \tilde K^{\rm M} =  - \log \int_{Y_4} \Omega \wedge \bar \Omega + \log \Big(\frac{1}{4!} \cK_{ABCD}L^A L^B L^C L^D \Big) +    \, L^A d_A{}^{l  k}  \ \R\, N_l  \ \R\, N_k \ ,
\eeq
and was already given in \eqref{def-tildeKM}. 
Recalling the discussion at the end of \autoref{22dilatonSugra} it is not hard to check 
that \eqref{3daction} reduces to the Type IIA result found 
in \autoref{IIAreduction_details} upon a circle compactification. 
The detailed circle reduction is performed for a general three-dimensional 
un-gauged $\cN=2$ theory in \autoref{3d-2dreduction}.

\subsection{M-theory to F-theory lift} \label{MFlimit}

Let us now lift the result \eqref{chirallinear3D} of the M-theory reduction on a general smooth Calabi-Yau fourfold $ Y_4 $ to a four-dimensional effective F-theory compactification. To do so, we need to restrict $ Y_4 $ to be an elliptic fibration $ \pi: Y_4 \rightarrow B_3 $ over a base manifold $ B_3 $ which is a three-dimensional complex K\"ahler manifold. This four-dimensional theory exhibits $ \cN=1 $ supersymmetry. 
In the following we will not need to consider the full four-dimensional theory, but 
will rather  focus on the kinetic terms of the complex scalars and vectors without 
including gaugings or a scalar potential. 
Supersymmetry ensures that these can be written in the form \cite{Wess:1992cp}
\beq \label{S4gen}
S^{(4)} =  \int \frac{1}{2} R \ast 1 
     - K^{\rm F}_{M^I \bar M^J} \, d M^I \wedge \ast d\bar{M}^J - \frac{1}{2} \R \, \mathbf{f}_{\Lambda \Sigma}
      F^\Lambda \wedge \ast F^\Sigma - \frac{1}{2} \I \,\mathbf{f}_{\Lambda \Sigma} F^\Lambda \wedge F^\Sigma\ .
\eeq
In this expression we denoted by $M^I$ the bosonic degrees of freedom in chiral multiplets,
and by $F^\Lambda$ the field strengths of vectors $A^\Lambda$. 
The metric $K^{\rm F}_{M^I \bar M^J} $ is K\"ahler and thus can be obtained from 
a K\"ahler potential $K^{\rm F}$ via $K^{\rm F}_{M^I \bar M^J} =  \partial_{M^I} \partial_{\bar M^J} K^{\rm F}$.
The gauge-kinetic coupling function $\mathbf{f}_{\Lambda \Sigma} $ is holomorphic 
in the complex scalars $M^I$. 

In order to determine the K\"ahler potential $K^{\rm F}$ and the gauge coupling 
function  $\mathbf{f}_{\Lambda \Sigma} $  via M-theory one next would have 
to compactify \eqref{S4gen} on a circle. The resulting 
three-dimensional theory then has to be pushed to the Coulomb 
branch and all massive modes, including the excited 
Kaluza-Klein modes of all four-dimensional fields, have to 
be integrated out. The resulting three-dimensional effective theory 
can then, after a number of dualizations, be compared with 
the M-theory effective action \eqref{3daction}.  
Performing all these steps is in general complicated. 
However, a relevant special case
has been considered in \cite{Grimm:2010ks} and 
will be the focus in the following discussion. 
Despite the fact that we could refer 
to \cite{Grimm:2010ks} we will try to keep the derivation of 
$K^{\rm F}$ and $\mathbf{f}_{\Lambda \Sigma} $ 
in this subsection self-contained.

Let us therefore assume that $Y_4$ is an elliptically fibered 
Calabi-Yau fourfold that satisfies the conditions
\beq  \label{special_geom}
 h^{2,1}(Y_4)= h^{2,1}(B_3)\ , \qquad \quad h^{1,1}(Y_4) = h^{1,1}(B_3) + 1\ .
\eeq
It is not hard to use toric geometry to construct examples
that satisfy these conditions (see, for example, refs.~\cite{Grimm:2009yu}). 
From the point of view of F-theory, or Type IIB string theory, the first 
condition in \eqref{special_geom} implies 
that all scalars $N_l$ in \eqref{3daction} lift to R-R vectors $A^l$ in 
four dimensions. In other words, one can compactify 
Type IIB on the base $B_3$ and obtain vectors $A^l$
by expanding the R-R four-form as 
\beq \label{C4expand}
     C_4 = A^l \wedge \alpha_I - \tilde A_l \wedge \beta^l + \ldots\ .
\eeq 
The vectors $\tilde A_l$ are the magnetic duals of the $A^l$
and can be eliminated by using the self-duality of the field-strength 
of $C_4$. 

The second condition in \eqref{special_geom}
implies that there are no further vectors in the four-dimensional 
theory, i.e.~there are no massless vector degrees of 
freedom arising from seven-branes. 
The two-forms used in \eqref{JM} and \eqref{11Dthree-form} split simply as
\beq
   \omega_A = (\omega_0, \omega_\alpha)\ ,
\eeq
where $ \omega_0$ is the Poincar\'e-dual of the base divisor $B_3$
and $\omega_\alpha$ is the Poincar\'e-dual of 
the vertical divisors $ D^\alpha = \pi^{-1} (D^\alpha_{\rm b}) $ stemming from 
divisors $D^{\alpha}_{\rm b}$ of  $B_3$. 
Accordingly one 
splits the three-dimensional vector multiplets in \eqref{chirallinear3D} 
as
\beq
    L^A = (R, L^\alpha )\ , \qquad V^A = (A^0, A^\alpha) \ . 
\eeq
One can now evaluate the kinetic potential \eqref{def-tildeKM_Recall}
for the special case \eqref{special_geom}. 
The only relevant non-vanishing quadruple 
intersection numbers are given by 
\beq \label{base-triple}
   \cK_{0\alpha \beta \gamma} = \int_{Y_4} \omega_0 \wedge \omega_\alpha \wedge \omega_\beta \wedge \omega_\gamma \equiv \cK_{\alpha \beta \gamma}\ , 
\eeq
which are simply the triple intersections $\cK_{\alpha \beta \gamma}$ of the base $B_3$. Crucially, 
for an elliptic fibration one has $\cK_{\alpha \beta \gamma \delta} = 0$. 
Furthermore, note that due to \eqref{special_geom} all non-trivial 
three-forms come from the base $B_3$ and we can chose the 
basis $(\alpha_I , \beta^l)$ such that 
\beq \label{special_C}
   C_{0m}{}^k = \int_{Y_4} \omega_0 \wedge \alpha_l \wedge \beta^k = \delta_l^k\ , \qquad C_{\alpha m}{}^{k} = C_{Amk} = 0\ , 
\eeq
with $C_{Am}{}^k$ and $C_{Amk}$ introduced in \eqref{Cdef}.
Inserting \eqref{base-triple} and \eqref{special_C} into \eqref{def-tildeKM_Recall} one finds
\beq \label{tildeKM_Special}
     \tilde K^{\rm M} =  - \log \int_{Y_4} \Omega \wedge \bar \Omega 
         + \log \Big(\frac{1}{3!} \cK_{\alpha \beta \gamma} L^\alpha L^\beta L^\gamma \Big) + \log(R)  - \frac{1}{2} R\, \R f^{l  k}  \ \R\, N_l  \ \R\, N_k \ ,
\eeq
where we have used that $L^A d_A{}^{l  k} =-\frac{1}{2} L^A C_{A m}^{l} \R f^{m k} =-\frac{1}{2} R\, \R f^{l k}$,
and we have dropped terms in the logarithm that are higher order in $R$.

In order to compare this kinetic potential with the result of the 
circle reduction of \eqref{S4gen} we next have to dualize $(L^\alpha, A^\alpha)$
into three-dimensional complex scalars $T_{\alpha}$, and $N_k$ into three-dimensional 
vectors $(\xi^k, A^k)$. Due to our assumption \eqref{special_geom}
leading to \eqref{special_C} we can perform these dualizations independently.
The change from $(L^\alpha,A^\alpha)$ to $\R T_\alpha = \partial_{L^\alpha} \tilde K^{\rm M}$
is similar to \eqref{TAbfK}. It is  conveniently 
parameterized by the base K\"ahler deformations $v^\alpha_{\rm b}$
and the base volume $\cV_{\rm b}$ defined as \cite{Grimm:2004uq,Grimm:2010ks}
\beq \label{def-vb}
     L^\alpha = \frac{v^\alpha_{\rm b}}{\cV_{\rm b}} \ , \qquad 
     \cV_{\rm b} = \frac{1}{3!} \cK_{\alpha \beta \gamma} v^{\alpha}_{\rm b} v^{\beta}_{\rm b} v^{\gamma}_{\rm b}\ .
\eeq 
The dualization of the complex scalars $N_k$ into three-dimensional vectors 
is similar to the dualization yielding \eqref{N'}, \eqref{tildeK'} and \eqref{N'ex}, \eqref{tildeK'ex}. First, one introduces 
\beq
\xi^k = \partial_{\R N_k} \tilde K^{\rm M} \ , \qquad  \tilde K^{\rm M \rightarrow F}  = \tilde K^{\rm M} -   \xi^k \R \, N_l \ ,
\eeq 
and then dualizes the 
field $\I N_k$ with a shift symmetry  into the vector $A^k$.   
Together both Legendre transforms yield
\beq \label{tildeKFM}
   \tilde K^{\rm M \rightarrow F} =  - \log \int_{Y_4} \Omega \wedge \bar \Omega 
         - 2 \log \, \cV_{\rm b}\, + \log \, R   + \frac{1}{2 R}\, \R f_{l  k}  \ \xi^l  \xi^k \ ,
\eeq
which has to be evaluated as a function of $z^K$, $\xi^k$ and 
\beq   \label{Talphabase}
 T_{\alpha} 
      = \partial_{L^\alpha} \tilde K^{\rm M}  + i \rho_\alpha
         =  \frac{1}{2!} \cK_{\alpha \beta \gamma}  v^\beta_{\rm b} v^\gamma_{\rm b}  + i \rho_\alpha\ .
\eeq
The kinetic potential \eqref{tildeKFM} is now in the correct frame to be lifted to 
four space-time dimensions. 

To derive  $K^{\rm F}$, $\mathbf{f}_{kl} $ 
one reduces \eqref{S4gen} on a circle of radius $r$ 
with the usual Kaluza-Klein ansatz 
the four-dimensional metric and vectors as
\beq
g^{(4)}_{\mu \nu} =
\begin{pmatrix}
 g^{(3)} _{pq} + r^2 A^0 _p A^0 _q & r^2 A^0 _q \\
 r^2 A^0 _p & r^2
\end{pmatrix},
\qquad A^k _\mu = (A^k_p + A^0 _p \zeta^k, \zeta^k)\ ,
\eeq
where we introduced the three-dimensional indices $ p,q = 0,1,2 $ and the Kaluza-Klein vector $ A^0 $. Note that we use for three-dimensional vectors the same symbol $ A^k $ as in four dimensions. 
Furthermore, we introduced the new three-dimensional real scalars $ r, \zeta^k $ into the theory. 
We next define 
\beq
R = r^{-2}\ , \quad \xi^{\hat k} = ( R, R \zeta^k)\ , \quad A^{\hat k} = (A^0, A^k) \ .
\eeq
The three-dimensional theory obtained by reducing \eqref{S4gen} has thus 
the field content: chiral multiplets with complex scalars $M^I$ and vector multiplets 
$(\xi^{\hat k},A^{\hat k} )$.
Its action can be written in the form \eqref{chirallinear3D} 
with a kinetic potential 
\beq \label{redK}
 \tilde K(M,\bar{M}, \xi) = K^F (M,\bar{M}) + \log (R) - \frac{1}{R} \R\, \mathbf{f}_{k l }(M) \xi^k \xi^l \  ,
\eeq
when replacing $L^A \rightarrow \xi^{\hat k}$, $V^A \rightarrow A^{\hat  k}$, and $\phi^\kappa \rightarrow M^I$.
Finally, comparing \eqref{redK} with \eqref{tildeKFM} implies that one finds $M^I = \{T_\alpha, z^K \} $
\bea
K^{\rm F} &=& -\log (\int_{Y_4} \Omega \wedge \overline{\Omega}) - 2 \log{\cV_b}\ , \label{F-theoryKpot}\\
 \mathbf{f}_{kl}&=& \frac{1}{2} f_{kl}\, .
\eea
In the next section, we want to derive the orientifold limit of this result relating the data of F-theory on $ Y_4 $ to Type IIB supergravity with $ O7/O3 $-planes on the closely related Calabi-Yau three-fold $ Y_3 $, a double cover of $ B_3 $. 

\subsection{Orientifold limit of F-theory and mirror symmetry} \label{orientifold_limit}
  
In this final subsection we investigate the orientifold limit of the F-theory
effective action introduced above. 
More precisely, we assume that the F-theory compactification on the 
elliptically fibered geometry $Y_4$ admits a weak string coupling limit as 
introduced by Sen \cite{Sen:1996vd,Sen:1997gv}. This limit takes one to a 
special region in the complex structure moduli space of $Y_4$
in which the dilaton-axion $\tau = C_0 + ie^{-\phi_{\rm IIB}}$, given by the complex 
structure of the two-torus fiber of $Y_4$, is almost everywhere 
constant along the base $B_3$. The locations where $\tau$
is not constant are precisely the orientifold seven-planes (O7-planes).  
In the weak string coupling limit the geometry $Y_4$ can 
be approximated by 
\beq \label{orientifold_projection}
Y_4 \cong (Y_3 \times T^2)/\tilde {\sigma}
\eeq
where we introduced the involution $ \tilde {\sigma} = (\sigma, -1,-1) $ with $ \sigma $ being a holomorphic and isometric 
orientifold involution such that $ Y_3 / \sigma = B_3 $. The two one-cycles of the torus are both odd under the involution, but its volume form is even. 
It was shown in \cite{Sen:1996vd,Sen:1997gv} that the double cover $Y_3$ of $B_3$ is actually a Calabi-Yau threefold. The location of the 
O7-planes in $Y_3$ is simply the fixed-point set of $\sigma$. 

In the limit \eqref{orientifold_projection} we can check compatibility of the mirror symmetry results of 
\autoref{mirror_section} with the mirror symmetry of the Calabi-Yau threefold $Y_3$.
By using the mirror fourfold $\hat Y_4$ of $Y_4$ we have found that the function $f_{lk}$ is linear in the large 
complex structure limit of $Y_4$. Here we recall that the weak string coupling expression gives a compatible result. Using 
the mirror $\hat Y_3$ of $Y_3$ one shows that the function $f_{lk}$ is linear in the large 
complex structure limit of $Y_3$. This can be depicted as
\beq
\begin{array}{ccc}
  \text{F-theory on}\ Y_4  &\quad  \xrightarrow{\quad \text{weak coupling} \quad } &  \text{Type IIB orientifolds}\ Y_3 / \sigma \qquad \\
  && \qquad\qquad\qquad \updownarrow \quad \text{physical mirror duality}\\
  &&\qquad \text{Type IIA orientifolds}\ \hat Y_3 / \hat \sigma \qquad
\end{array}
\eeq
Note that mirror symmetry of $Y_3$ and $\hat Y_3$ gives a physical map between Type IIB and Type IIA orientifolds.
The mirror map between $Y_4$ and $\hat Y_4$ has no apparent physical meaning in F-theory. Nevertheless, 
using the geometry $Y_4$ in Type IIA compactifications it can be used to calculate 
$f_{lk}$ as we explained in \autoref{mirror_section}. 

Let us now introduce the function $f_{lm}$ for the geometry \eqref{orientifold_projection}. 
In the orientifold setting one splits the cohomologies of $Y_3$ as $H^{p,q} (Y_3)= H^{p,q}_+ (Y_3) \oplus H^{p,q}_- (Y_3)$,
which are the two eigenspaces of $\sigma^*$. We denote their dimensions as $h^{p,q}_\pm (Y_3)$.
As reviewed, for example, in \cite{Denef:2008wq} 
the complex structure moduli $z^K$ of $Y_4$ split into three sets of fields at weak string coupling.
First, there is the dilaton-axion $\tau$, which is now a modulus of the effective theory. 
Second, there are $h^{2,1}_-$ complex structure moduli $z^{\kappa}$  of the quotient $Y_3/\sigma$. 
Third, the remaining number of complex structure deformations of $Y_4$  correspond to 
D7-brane position moduli. The last set are open string degrees of freedom
and are not captured by the geometry of $Y_3$. For simplicity, we will not include 
them in the following discussion. 
With this simplifying assumption  one finds that the pure complex structure part of the F-theory 
K\"ahler potential \eqref{F-theoryKpot} splits as
\beq \label{Omegasplit}
-\log (\int_{Y_4} \Omega \wedge \overline{\Omega} ) = - \log \big[ -i(\tau - \bar{\tau})\big] - \log \Big[ i \int_{Y_3} \Omega_3 \wedge \bar{\Omega}_3 \Big] +\ldots \, ,
\eeq
where $\Omega_3$ is the $(3,0)$-form on $Y_3$ that varies holomorphically 
in the complex structure moduli $z^{\kappa}$.  
The dots indicate that further corrections arise that are suppressed at weak string coupling $-i(\tau - \bar \tau) \gg 1$. 
Taking the weak coupling limit for the K\"ahler potential \eqref{F-theoryKpot} of the 
K\"ahler structure deformations is more straightforward. The deformations are counted by 
$h^{1,1}_+(Y_3)$ and identified with the K\"ahler structure deformations $v_{\rm b}^\alpha$ of the 
base $B_3$ introduced in \eqref{def-vb}. The orientifold K\"ahler potential for this 
set of deformations is then simply the second term in \eqref{F-theoryKpot} and 
the K\"ahler coordinates are given by \eqref{Talphabase}.

Turning to the gauge theory sector, we note that the number of R-R vectors $A^l$ arising 
from $C_4$ as in \eqref{C4expand} are counted by $h^{2,1}_+(Y_3)$ in the orientifold setting. 
The gauge coupling function for these vectors is determined as function of 
the complex structure moduli $z^{\kappa}$ of $Y_3$ in \cite{Grimm:2004uq}.\footnote{Note that we 
have slightly changed the index conventions with respect to \cite{Grimm:2004uq} in order to 
match the F-theory discussion. } 
It is given by 
\beq \label{f-ori}
  f_{kl}(z^\kappa) = -i \cF_{k l}|(z^{\kappa}) \equiv \partial_{z^k} \partial_{z^l} \cF | (z^{\kappa})\ ,
\eeq
where $\cF$ is the pre-potential determining the moduli-dependence of the 
$\Omega_3$  of the geometry $Y_3$. To evaluate \eqref{f-ori} one first 
splits the complex structure moduli of $Y_3$ into $h^{2,1}_-(Y_3)$ fields 
 $z^{\kappa}$ and $h^{2,1}_+(Y_3)$ fields $z^k$. The pre-potential $\cF(z^{\kappa},z^k)$
 of $Y_3$ at first depends on both sets of fields. 
 Then one has to take derivatives of $\cF$ with respect to $z^k$ and 
 afterwards set these fields to constant background values compatible 
 with the orientifold involution $\sigma$. This freezing of the $z^k$ is indicated 
 by the symbol $|$ in \eqref{f-ori}. 
 Using mirror symmetry for Calabi-Yau threefolds it is well-known 
 that the pre-potential at the large complex structure point of $Y_3$ 
is a cubic function of the complex structure moduli $z^{\kappa}$ and $z^{k}$.
Taking derivatives and evaluating the expression on the orientifold moduli 
space one thus finds
\beq
f_{k l}(z^\kappa) =  -i z^{\kappa} \hat \cK_{\kappa k l } \ ,
\eeq
where $ \hat \cK_{\kappa k l } = \int_{\hat{Y}_3} \hat{\omega}_{\kappa} \wedge \hat{\omega}_k \wedge \hat{\omega}_l $
are the triple intersection numbers of the mirror threefold $\hat Y_3$.
This result agrees with the one for Type IIA orientifolds, which have 
been studied at large volume in \cite{Grimm:2004ua}.
Hence, we find consistency with the F-theory result \eqref{lin_fresults} obtained by using 
mirror symmetry for $Y_4$ at the large complex structure point. To obtain a
complete match of the results the 
intersection matrix $\hat C_{\kappa k}{}^l$ of $\hat Y_4$ is identified with the triple intersection $\hat \cK_{\kappa k l } $
of $\hat Y_3$. 

To close this section we stress again that we have only discussed the matching with 
the orientifold limit for special geometries satisfying \eqref{special_geom}. 
Furthermore, we have not included the open string degrees of freedom on the 
orientifold side. Clearly, our result for $f_{lk}$ obtained in \autoref{mirror_section} 
can be more generally applied. For example, a simple generalization 
is the inclusion of $h^{1,1}_-(Y_3)$ moduli $G^a$ into the orientifold setting, 
which arise in the expansion of the complex two-form $C_2-\tau B_2$. 
In F-theory the same degrees of freedom appear
from the expansion \eqref{11Dthree-form} into non-trivial three-forms $\Psi_a$ that have two legs 
in the base $B_3$ and one leg in the torus fiber, i.e.~are not present in 
the geometries satisfying \eqref{special_geom}.
In the orientifold setting one finds that the fields $G^a$ correct the complex 
coordinates \eqref{Talphabase}. We read off the result from \cite{Grimm:2004uq} to find \footnote{Note that 
compared with \cite{Grimm:2004uq} we have redefined $\rho_\alpha$ to make the terms
in $T_\alpha$ involving the $G^a$ real.}
\beq
T_\alpha =\frac{1}{2!} \cK_{\alpha \beta \gamma}  v^\beta_{\rm b} v^\gamma_{\rm b}  
    +  \frac{1}{2 \, \I \tau}\ \cK_{\alpha a b}\ \I \, G^a \I \, G^b  + i \rho_\alpha \, .
 \eeq
Comparing this expression with \eqref{TAbfK} we read off that 
\beq
   N^a = i G^a\ , \qquad d_{\alpha ab} = \frac{1}{2} \frac{1}{ \I \tau} \cK_{\alpha a b}\ , \qquad f_{ab} (\tau)= i \tau \delta_{ab}\ ,
\eeq
in order to match the F-theory result as already done in \cite{Grimm:2005fa}. 
Again we find that the result is linear in one of the complex structure moduli, namely the field $\tau$, 
of the Calabi-Yau fourfold $Y_4$ in the orientifold limit \eqref{orientifold_projection}. 
It would be interesting to generalize these results even further and also include the open 
string moduli into the orientifold setting.
 
\section{Conclusions}

In this paper we first studied 
the two-dimensional low-energy effective action obtained from Type IIA string theory 
on a Calabi-Yau fourfold with non-trivial 
three-form cohomology. 
The couplings of the three-forms were shown to 
be encoded by two holomorphic functions $f_{kl}$ and 
$h_k^l$, where the former depends on the complex structure moduli 
and the latter on the complexified K\"ahler structure moduli. 
Performing a large volume dimensional reduction of Type IIA supergravity, we 
were able to derive $h_k^l$ explicitly as a linear function. 
We argued that $f_{kl}$ and $h_k^l$ computed 
on mirror pairs of Calabi-Yau manifolds will be exchanged, at least,
if one considers the theories at large volume and large complex structure.
In order to show this, we investigated the non-trivial map 
between the three-form moduli arising from mirror geometries 
and argued that it involves a scalar field dualization together 
with a Legendre transformation.
This can be also motivated by the fact that chiral and 
twisted-chiral multiplets are expected to be exchanged 
by mirror symmetry. We thus established a linear dependence of  
the function $f_{lk}$ on the complex structure moduli near 
the large complex structure point and determined the constant topological
pre-factor. 

In this work we also included a  discussion of 
the superymmetry properties of the two-dimensional 
low-energy effective action. This action is expected to be an 
$\cN=(2,2)$ supergravity theory, which we showed to 
extend the dilaton supergravity action of \cite{Gates:2000fj}.
The bosonic action was brought to an elegant 
form with all kinetic and topological terms determined 
by derivatives of a single function $\tilde K = \cK + e^{2 \tilde \varphi} \cS$,
where $\cK$ and $\cS$ can depend on the scalars in chiral and 
twisted-chiral multiplets, but are independent of the two-dimensional  dilaton $\tilde \varphi$. 
In the Type IIA supergravity reduction the three-form scalars only 
appeared in the function $\cS$ and are thus suppressed by 
$e^{2 \tilde \varphi} = e^{2 \phi_{\rm IIA}}$. 
In this analysis the complex structure moduli and the 
three-form moduli were argued to fall into chiral multiplets, while the  
complexified K\"ahler moduli are in twisted-chiral multiplets. 
However, due to apparent shift symmetries of the three-form moduli 
and complexified K\"ahler moduli a scalar dualization accompanied by 
a Legendre transformation can be performed in 
two dimensions. This lead to dual descriptions in which certain 
chiral multiplets are replaced by twisted-chiral multiplets and vice versa. 
Remarkably, if one dualizes a subset of scalars appearing in $\cK$, 
we found that the requirement 
to bring the dual action back to the standard $\cN=(2,2)$ dilaton supergravity 
form imposes conditions on viable $\cK$.
These constraints include a no-scale type condition on $\cK$. 
The emergence of such restrictions arose from general arguments 
about two-dimensional theories coupled to an overall
$e^{-2\tilde \varphi}$ factor. For Calabi-Yau fourfold reductions we checked that these conditions are
indeed satisfied. 
It would be interesting to investigate this further and to get a  
deeper understanding of this result.

Having shown that in the large complex structure limit
the function $f_{kl}$ is linear in the complex structure 
moduli, we discussed the 
application of this result in an F-theory compactification. 
By assuming that the Calabi-Yau fourfold is elliptically fibered and that
the three-forms exclusively arise from the base of this fibration, 
we recalled that $f_{kl}$ is actually the gauge-coupling 
function of four-dimensional R-R vector fields. 
This gauge-coupling function was already evaluated 
in the weak string coupling limit in the orientifold literature. 
In this orientifold limit 
one can double-cover the base with a Calabi-Yau threefold. 
We found compatibility of the fourfold result with the expectation
from mirror symmetry for Calabi-Yau threefold orientifolds. 
In this analysis we only included closed string moduli in 
the orientifold setting. Clearly, the results obtained from the 
Calabi-Yau fourfold analysis are more powerful and it would be 
interesting to further investigate the open string dependence in 
orientifolds using our results. Additionally we commented briefly 
on the case in which the three-forms have legs in the fiber 
of the elliptic fibration. In this situation the inverse of $\R f_{lk}$
sets the value of decay constants of four-dimensional axions \cite{Grimm:2014vva}. 
Again we found compatibility in the closed string sector at weak string 
coupling in which $f_{lk} \propto i \tau$. It would be interesting 
to include the open string moduli in the orientifold 
setting and derive corrections to $f_{lk}$ without restricting to the 
weak string coupling limit. The latter task requires to compute $f_{lk}$ away from the 
large complex structure limit for elliptically fibered Calabi-Yau fourfolds.

In order to derive the complete moduli dependence of $f_{lk}$ 
at various points in complex structure moduli space it would 
be desirable to obtain differential equations obeyed by 
the $(2,1)$-forms. This should be possible by investigating 
the variations of Hodge structures and is expected to yield 
equations of second order in derivatives. The linear solutions 
found in this work can then provide the boundary 
conditions for the complete solutions. It would be important to 
develop the necessary tools for such an analysis and 
we hope to return to this issue in a future publication.

 \subsubsection*{Acknowledgments}
We would like to thank Hans Jockers, Andreas Kapfer, Denis Klevers,  
Diego Regalado, and Matthias Weissenbacher for illuminating discussions. 
This work was supported by a grant of the Max Planck Society.

\appendix

\section*{Appendices}

\section{Three-dimensional $ \cN=2 $ supergravity on a circle}  \label{3d-2dreduction}

In this appendix we consider $\cN=2$ supergravity compactified on 
a circle of radius $r$. Our goal is to derive the resulting $\cN=(2,2)$ 
action. We also briefly discuss the dualization of vector multiplets in 
three dimensions and point out the relation to \autoref{detailed_dual}.
  
We start with a three-dimensional $ \cN=2 $ supergravity theory 
coupled to chiral multiplets with complex scalars $ \phi^\kappa $ and vector
multiplets with bosonic fields $ (L^A, A^A) $. Here$ L^A $ is a real scalar and $ A^A $ a vector of
an $ U(1) $ gauge theory. The bosonic part of the ungauged $\cN=2$ action takes the form
\begin{align} \label{3Dsugra}
S^{(3)} &= \int \frac{1}{2} R^{(3)} \ast 1 
      - \tilde{K}_{\phi^\kappa \bar \phi^\lambda} d\phi^\kappa \wedge \ast d\bar{\phi}^\lambda  
          + \frac{1}{4} \tilde{K}_{L^A L^B} dL^A \wedge \ast dL^B \nonumber \\
	&\qquad + \frac{1}{4} \tilde{K}_{L^A L^B} \ dA^A \wedge \ast dA^B + dA^A \wedge \I(\tilde{K}_{L^A \phi^\kappa} d\phi^\kappa) 
\end{align}
where the kinetic terms of the vectors and scalars are determined by the single real kinetic potential $\tilde{K}$.

We want to put this on a circle of radius $ r $ anvd period one, i.e. the background metric is of the form
\beq \label{circle_ansatz}
ds^2_{(3)} = g_{\mu \nu} dx^\mu dx^\nu + r^2 dz^2
\eeq
where we already drop vectors, since in an un-gauged theory they do not carry degrees of freedom
 in two dimensions. Similarly, the vectors $ A^A $ are only reduced to real scalars $dA^A =  d b^A \wedge dz$.
The resulting two-dimensional action thus reads
\begin{align} \label{actcc}
S^{(2)} &= \int \frac{1}{2} rR  \ast 1 - r\tilde{K}_{\phi^\kappa \bar \phi^\lambda } d\phi^\kappa \wedge \ast d\bar{\phi}^\lambda  
       + \frac{1}{4} r \tilde{K}_{L^A L^B} dL^A \wedge \ast dL^B \nonumber \\
	&\qquad \quad+ \frac{1}{4 r} \tilde{K}_{L^A L^B} db^A \wedge \ast db^B 
	- db^A \wedge \I (\tilde{K}_{L^A \phi^\kappa} d\phi^\kappa)\ , 
\end{align}
with a two-dimensional $R$ and Hodge star $*$. Note that the last term is topological and does 
not couple to the radius $ r $ of the circle.
We can perform  Weyl rescaling of the two-dimensional  metric setting  $\tilde g_{\mu \nu} = e^{2 \omega} g_{\mu \nu}$.
This transforms the Einstein-Hilbert term as 
\beq \label{WeylEH}
   \int   \frac{1}{2} r \tilde R \ \tilde *1    = 
    \int \frac{1}{2} r\, R \  *1 + d\omega \wedge * dr \ ,
\eeq
while leaving all other terms in the action \eqref{actcc} invariant. 
We then find the action
\begin{align}  \label{action_new1}
S^{(2)} &= \int r \left( \frac{1}{2} R \ast 1 + d\log r \wedge \ast d\omega - \tilde{K}_{\phi^\kappa\bar \phi^\lambda } d\phi^\kappa \wedge \ast d\bar{\phi}^\lambda  + \frac{1}{4} \tilde{K}_{L^A L^B} dL^A \wedge \ast dL^B \right.  \nonumber \\
	&\qquad \qquad\left. + \frac{1}{4r^2} \tilde{K}_{L^A L^B} db^A \wedge \ast db^B \right) - db^A \wedge \I(\tilde{K}_{L^A \phi^\kappa} d\phi^\kappa) 
\end{align}
To make contact with the $ \cN=(2,2) $ dilaton supergravity action \eqref{bosonic_dilatonsugra_extended}
we set
\bea
 L^A  &=&  r^{-1}  v^A\ ,\qquad  r= e^{-2\tilde{\varphi}}  \ ,\\
 \sigma^A &\equiv& b^A + i v^A\ .
\eea
Inserted into \eqref{action_new1} we then obtain
\begin{align} \label{action_new2}
S^{(2)} &= \int e^{-2 \tilde \varphi} \left( \frac{1}{2} R  \ast 1 - 2 d \tilde \varphi \wedge \ast 
        \Big(d\omega - \frac{1}{2}  \tilde{K}_{v^A v^B}v^A dv^B - \frac{1}{2}\tilde{K}_{v^A v^B} v^A  v^B d\tilde \varphi \Big)   \right.  \\
&\qquad \qquad \qquad \left.- \tilde{K}_{\phi^\kappa\bar \phi^\lambda } d\phi^\kappa \wedge \ast d\bar{\phi}^\lambda 
 +  \tilde{K}_{\sigma^A \bar \sigma^B} d\sigma^A \wedge \ast d \bar \sigma^B  - d\R\, \sigma^A \wedge \I(\tilde{K}_{v^A \phi^\kappa} d\phi^\kappa) \right)  \ .  \nonumber 
\end{align}
In order to match the action \eqref{bosonic_dilatonsugra_extended} one 
therefore has to find an $\omega$ such that
\beq \label{dw=phi}
      d\omega = -d \tilde \varphi + \frac{1}{2}  \tilde{K}_{v^A v^B}v^A dv^B + \frac{1}{2}\tilde{K}_{v^A v^B} v^A  v^B d\tilde \varphi 
\eeq
To solve this condition, we first notice that any term in $\tilde K$ that is linear $v^A$ drops out from this 
relation, i.e.~$\tilde K$ can take the form
\beq \label{SAsplit}
   \tilde K = \cK + v^A \cS_A\ ,
\eeq
with an arbitrary function $\cS_A(\phi,\bar \phi)$. Furthermore, we can solve \eqref{dw=phi} by 
assuming that $\cK= \cK_1 + \cK_2$ splits into a $v^A$-independent term $\cK_1(\phi,\bar \phi)$ and a
term $\cK_2(v)$ that only depends on $v^A$. Then \eqref{dw=phi} is satisfied if
\beq \label{no-scaleappM}
  v^A \cK_{v^A} = -k \ , \qquad \omega = - \tilde \varphi + \frac{k}{2} \tilde \varphi -\frac{\cK_2(v)}{2} \ , 
\eeq
It is easy to check that the conditions \eqref{SAsplit} and
\eqref{no-scaleappM} are actually satisfied for the M-theory example \eqref{def-tildeKM}
of $\tilde K$. One finds 
\beq
   \cK_1(z) = - \log \int_{Y_4} \Omega \wedge \bar \Omega \ ,\qquad   \cK_2(v) = \log \cV\ , \qquad 
   \cS_A= e^{2\varphi} d_A{}^{l \bar k}\  \R N_l\  \R N_k\ ,
\eeq
such that $k = -4$. Finally, in order to show that \eqref{action_new2} is indeed identical
to the action \eqref{bosonic_dilatonsugra_extended}, we still have to complete the last term
in \eqref{action_new2} to $\I (d \sigma^A \wedge \tilde{K}_{v^A \phi^\kappa} d\phi^\kappa)$. 
In order to do that we  use 
\beq
    d\I \, \sigma^A \wedge \R(\tilde{K}_{v^A \phi^\kappa} d\phi^\kappa)=    \frac{1}{2} d\I \, \sigma^A \wedge d \tilde K_{v^A} \ ,
\eeq
which follows from the fact that $d \tilde K_{v^A} = 2 \R( \tilde K_{v^A \phi^\kappa} d \phi^\kappa )  + \tilde K_{v^A v^B} d v^B$.
This implies that these terms simply yield a total derivative 
and shows that the reduction of $\cN=2$ supergravity of the form \eqref{3Dsugra} indeed 
yields the  extended form of 
$ \cN=(2,2) $  dilaton supergravity suggested in \eqref{bosonic_dilatonsugra_extended} 
coupled to the chiral multiplets with scalars $ \phi^\kappa $ and twisted-chiral multiplets with 
scalars $ \sigma^A $. Interestingly, we had to employ the conditions \eqref{SAsplit} and \eqref{no-scaleappM}, which hints
to the fact that the action \eqref{bosonic_dilatonsugra_extended} might admit further interesting
extensions. 

Let us end this appendix by pointing out that we could also have first dualized the vectors $ A^A $
to real scalars in three dimensions and then performed the circle reduction. 
The dual multiplets to the vector multiplets $ (L^A, A^A) $ 
are three-dimensional chiral multiplets with bosonic parts being complex scalars $ T_A $ 
given by
\beq \label{3Dchiral}
T_A =  \partial_{L^A} \tilde K  + i \rho_A\ .
\eeq
The metric is determined now from a proper K\"ahler potential given by 
\beq
\textbf{K}(T+\bar T, M) = K - \R\, T_A \, L^A \ ,
\eeq
such that the final action reads
\begin{align}   
S^{(3)} &= \int \frac{1}{2} R^{(3)} \ast 1 - \textbf{K}_{{M}^I \bar{ { M}}^J } d {M}^I \wedge \ast d\bar{{M}}^J
\end{align}
with $ {M}^I= (T_A, \phi^\kappa )$. We can again reduce this theory on a circle \eqref{circle_ansatz}
and perform a Weyl-rescaling \eqref{WeylEH} to find
\begin{align}
S^{(2)} &= \int \frac{1}{2} rR \ast 1 + dr \wedge \ast d\omega - r\textbf{K}_{{M}^I \bar{ { M}}^J } d{M}^I \wedge \ast d\bar{{M}}^J\ .
\end{align}
With the choices $ r = e^{-2\tilde{\varphi}}$ and $ \omega = - \tilde{\varphi} $ this reads
\begin{align}
S^{(2)} &= \int  e^{-2\tilde{\varphi}}\left(\frac{1}{2} R\ast 1 + 2d\tilde{\varphi} \wedge \ast d\tilde{\varphi} 
- \mathbf{K}_{{M}^I \bar{ { M}}^J } d{M}^I \wedge \ast d\bar{{M}}^J \right)\ .
\end{align}
This result should also be obtainable from \eqref{action_new2} by dualizing the chiral multiplets
with scalars $\sigma^A$. This is possible since $ b^A $ appears only with its field-strength $db^A $. 
The details of this dualization in two dimensions will be discussed in \autoref{detailed_dual}.

\section{Twisted-chiral to chiral dualization in two dimensions} \label{detailed_dual}

In this appendix we present the details of the dualization discussed 
in \autoref{sec_Legendre} of 
a twisted-chiral multiplet to a chiral multiplet in two dimensions. 
The starting point is the action 
\begin{align}  \label{action1app}
 S^{(2)}_{\text{C-TC}} = \int e^{-2\tilde{\varphi}} & \left( \frac{1}{2}R *1 + 2d \tilde{\varphi} \wedge * d\tilde{\varphi} 
 -  \tilde K_{\phi^\kappa \bar \phi^\lambda} \, d\phi^\kappa \wedge * d \bar {\phi}^\lambda
	  +  \tilde  K_{ \sigma^A \bar \sigma^B}\, d \sigma^A \wedge * d \bar {\sigma}^B \right.\nonumber \\
   & - \left. \tilde K_{ \phi^\kappa \bar \sigma^B} \, d \phi^\kappa \wedge d \bar \sigma^B
	  - \tilde K_{ \sigma^A \bar \phi^\lambda} \, d \bar {\phi}^\lambda \wedge d\sigma^A \right)\ ,
\end{align}
where $\tilde K$ is given by 
\beq 
    \tilde K = \cK + e^{2 \tilde \varphi} \cS\ . 
\eeq
In the following we 
use sub-scripts to indicate derivatives with respect to fields, e.g.~$\tilde K_{\phi^\kappa} \equiv \partial_{ \phi^\kappa} \tilde K $.
$\tilde K$ depends on a number of chiral multiplets with complex
scalars $\phi^\kappa$ and a number of twisted-chiral 
multiplets with complex scalars $\sigma^A$.

In order to perform a dualization, we assume that 
$\R\, \sigma^A$ has a shift symmetry and only appears 
via $d \,\R\, \sigma^A$ in \eqref{action1app}. 
This implies that $\R\, \sigma^A$ can be dualized 
into a scalar $\rho_A$ by the standard procedure. One 
first replaces $d\R\, \sigma^A \rightarrow F^A$ in \eqref{action1app} 
and then adds a Lagrange multiplier term promotional to 
$F^A \wedge d\rho_A$. Then $F^A$ can be consistently eliminated 
from \eqref{action1app}. 
Denoting the imaginary part of $\sigma^A$ by $v^A = \I \, \sigma^A$
the resulting action reads
\begin{align}  \label{action2app}
 S^{(2)}_{\rm C} = \int e^{-2\tilde{\varphi}} & \left( \frac{1}{2}R *1 + 2d \tilde{\varphi} \wedge * d\tilde{\varphi}
 -  \tilde K_{\phi^\kappa \bar \phi^\lambda} \, d\phi^\kappa \wedge * d \bar {\phi}^\lambda
      + \frac{1}{4} \tilde  K_{ v^A  v^B}\, d v^A \wedge * d v^B \right.  \\
   & \left. + \tilde K ^{v^A v^B} \big(e^{2\tilde \varphi} d\rho_A - \I \, (\tilde K_{v^A \phi^\kappa} d\phi^\kappa) \big) \wedge 
        \ast \big(e^{2\tilde \varphi} d\rho_B - \I \, (\tilde K_{v^B \phi^\lambda} d\phi^\lambda) \big) \right) \nn
\end{align}

To compute the dualized action we make the following
ansatz for the Legendre transformed variables $T_A$ 
\beq \label{TA_app}
   T_A =  e^{-2\tilde \varphi} \frac{\partial \tilde K}{\partial v^A} + i \rho_A=
    e^{-2\tilde \varphi} \frac{\partial \cK}{\partial v^A}+ \frac{\partial \cS}{\partial v^A} + i \rho_A \ , 
\eeq
and the dual potential $\mathbf{K}$ 
\beq \label{bfK_app}
    \mathbf{K} = \tilde K - e^{2\tilde \varphi}\, \R \, T_A \, v^A\ .  
\eeq
We want to derive the conditions on $\tilde K$ under which the action \eqref{action2app} can be 
brought to the form
\beq \label{S2_app}
   S^{(2)}_{\text{C}} = \int e^{-2\tilde{\varphi}} \left( \frac{1}{2}R *1 + 2d \tilde{\varphi} \wedge * d\tilde{\varphi} 
 -  \mathbf{K}_{M^I \bar M^J} \, dM^I\wedge * d \bar M^J \right)\ ,
\eeq
with $M^I = (\phi^\kappa, T_A)$.

We first determine from \eqref{TA_app} and \eqref{bfK_app} that
\begin{align} \label{usefull}
   \frac{\partial v^A}{\partial T_B} &= \frac{1}{2} e^{2 \tilde \varphi} \tilde K^{v^A v^B} \ ,  
     &\frac{\partial v^A}{\partial \phi^\kappa} &= -  \tilde K^{v^A v^B} \tilde K_{v^B  \phi^\kappa}\ , \\
     \mathbf{K}_{T_A} &= - \frac{1}{2} e^{2 \tilde \varphi} v^A\ ,   & \mathbf{K}_{ \phi^\kappa} &= \tilde K_{ \phi^\kappa}\ , \nn
\end{align}
where $\tilde K^{v^A v^B}$ is the inverse of $\tilde K_{v^A v^B} \equiv \partial_{v^A} \partial_{v^B} \tilde K 
= 4 \tilde K_{\sigma^A \bar \sigma^B}$. Crucially, one also derives from \eqref{TA_app} that
\beq \label{dRT}
  d \R T_A = e^{-2 \tilde \varphi} \big( \tilde K_{v^A v^B} d v^A + 2 \R ( \tilde K_{v^A  \phi^\kappa} d \phi^\kappa ) - 2 \cK_{v^A} d\tilde \varphi \big)\ .
\eeq
Note that there is the additional $d\tilde \varphi $-term, which is absent in the standard dualization procedure. 
The conditions on $\tilde K$ arise from demanding that the dual action can be brought to 
the form \eqref{action1app} and no additional mixed terms involving $d\tilde \varphi $ appear. 
To evaluate \eqref{action1app} one uses \eqref{usefull} to derive the identities
\bea  \label{metric1}
  \mathbf{K}_{T_A \bar T_B} &=& - \frac{1}{4} e^{4 \tilde \varphi} \tilde K^{v^A v^B}\ , \qquad 
  \mathbf{K}_{T_A \bar \phi^\kappa} =  \frac{1}{2} e^{2 \tilde \varphi} \tilde K^{v^A v^B} \tilde K_{v^B \bar \phi^\kappa}\ , \\
  \mathbf{K}_{\phi^\kappa \bar \phi^\lambda} &=&   
  \tilde {K}_{\phi^\kappa \bar \phi^\lambda} -  \tilde K_{\phi^\kappa v^A}  \tilde K^{v^A v^B} \tilde K_{v^B \bar \phi^\lambda}\ .\nn
\eea
Inserting \eqref{dRT}, \eqref{metric1} into \eqref{S2_app} one finds the following terms involving $d \tilde \varphi$
\beq \label{extradphi}
  S_{d \tilde \varphi}^{(2)} = \int e^{-2 \tilde \varphi} \Big( \big(2 + \cK_{v^A}  \cK^{v^A v B} \cK_{v^B} \big)
  d\tilde \varphi \wedge * d \tilde \varphi + \cK_{v^A} dv^A \wedge * d \tilde \varphi \Big)\ .
\eeq
These terms can be removed by a Weyl rescaling of the three-dimensional metric 
if certain conditions on $\cK$ are satisfied. To see this, we perform a Weil rescaling 
\beq \label{Weyl-change}
      \tilde g_{\mu \nu} = e^{2 \omega} g_{\mu \nu}
\eeq
which transforms the Einstein-Hilbert term as 
\beq \label{change_R}
   \int  e^{-2 \tilde \varphi}  \frac{1}{2} \tilde R \ \tilde *1    = 
    \int e^{-2 \tilde \varphi}\left( \frac{1}{2} R \  *1 - 2 d\omega \wedge * d\tilde \varphi \right) \ ,
\eeq
while leaving all other terms invariant. Hence we can absorb the extra terms 
in \eqref{extradphi} by a Weyl rescaling iff
\beq
  - 2 d\omega = \cK_{v^A}  \cK^{v^A v B} \cK_{v^B} d\tilde \varphi + \cK_{v^A} dv^A\ .
\eeq
Clearly, a simple solution to this equation is found if $\cK$ satisfies 
\beq \label{nsc}
   \cK_{v^A}  \cK^{v^A v B} \cK_{v^B}  = k\ , \qquad  \cK = \cK_1(\phi,\bar \phi) + \cK_2(v)\ , 
\eeq
for a constant $k$, a function $\cK_1(\phi,\bar \phi) $ independent of $v^A$, and a function 
$\cK_2(v)$ independent of $\phi^\kappa$. 
In this case one can chose
\beq
    \omega = - \frac{k}{2} \tilde \varphi - \frac{1}{2} \cK_2(v)\ . 
\eeq
Note that \eqref{nsc} is satisfied for the result found in a Calabi-Yau fourfold reduction \eqref{idhatKS}, 
i.e.~$k=-4$ and $\cK_2 = \log \cV$.

\end{document}